\documentclass[manuscript]{acmart}

\usepackage{etoolbox}
\usepackage{expl3}
\usepackage{array}
\usepackage{booktabs}
\usepackage{calc}
\usepackage{euscript}
\usepackage{extarrows}
\usepackage[bottom,hang,flushmargin]{footmisc}
\usepackage{graphicx}
\usepackage{graphbox}
\usepackage{listings}
\usepackage{microtype}
\usepackage{multicol}
\usepackage{siunitx}

\usepackage{soul}
\setulcolor{black}
\setstcolor{black}
\sethlcolor{black}

\usepackage{subcaption}
\usepackage{textcomp}
\usepackage[absolute,overlay]{textpos} 
\TPoptions{showboxes=false}
\TPMargin*{2pt} 
\usepackage{upquote}
\usepackage{verbatim}

\usepackage{catchfilebetweentags}
\makeatletter
\patchcmd{\CatchFBT@Fin@l}{\endlinechar\m@ne}{}{}{}
\makeatother

\makeatletter
\@ifclassloaded{acmart}{}{
  \usepackage{amsmath,amssymb,amsthm}
}
\@ifclassloaded{beamer}{} 
{%
  \usepackage[inline,shortlabels]{enumitem}
  \setlist[itemize]{topsep=0pt,itemsep=0pt}

  \usepackage{todonotes}

  \usepackage{hyperref}
  \hypersetup{
    colorlinks=true,
    linkcolor=white,
    urlcolor=blue
    final,                   
    hidelinks,               
    breaklinks=true,         
    anchorcolor=blue,        
  }

  \newcommand{\inlinetodo}[1] {\todo[inline]{#1}\noindent}
  \newcommand{\reference}[1] {\footnotesize{\textbf{Reference:} #1}}
  \newcommand{\setlistspacing}[3] {}

  \newcommand{\showframe}[1] {%
    \begin{center}
      \framebox{\includeslide[width=8cm]{#1}}
    \end{center}
  }

  \newcommand*{\titleslide}[1] {}
}
\makeatother

\definecolor[named]{vlightgrey}{gray}{0.9}
\definecolor[named]{beamergreen}{rgb}{0.2,0.5,0.5}
\definecolor[named]{mygreen}{rgb}{0.4,0.9,0.7}
\definecolor[named]{myred}{rgb}{1,0.7,0.5}

\providecommand{\amber}{}
\providecommand{\blue}{}
\providecommand{\brown}{}
\providecommand{\green}{}
\providecommand{\grey}{}
\providecommand{\red}{}
\providecommand{\white}{}

\renewcommand{\amber}[1] {\textcolor[rgb]{1.00,0.50,0.00}{#1}}
\renewcommand{\blue}[1] {\textcolor{blue}{#1}}
\renewcommand{\brown}[1] {\textcolor{Mahogany}{#1}}
\renewcommand{\green}[1] {\textcolor[rgb]{0,0.7,0}{#1}}
\renewcommand{\grey}[1] {\textcolor{gray}{#1}}
\newcommand{\mygreen}[1] {\textcolor{mygreen}{#1}}
\newcommand{\myred}[1] {\textcolor{myred}{#1}}
\renewcommand{\red}[1] {\textcolor{red}{#1}}
\renewcommand{\white}[1] {\textcolor{white}{#1}}

\ifdefined\labelenumiii
  \renewcommand{\labelenumiii}{(\roman{enumiii})}
\else
  \newcommand{\labelenumiii}{(\roman{enumiii})}
\fi

\newfont{\helvetica}{phvr8t at 10pt}
\newfont{\helveticasmall}{phvr8t at 9pt}
\newfont{\helveticabig}{phvr8t at 22pt}
\newfont{\helveticao}{phvro8t at 11pt}

\newcommand{\smallsc}[1] {{\footnotesize\textsc{#1}}}
\newcommand{\setfontsizeandspacing}[2] {\fontsize{#1}{#2}\selectfont}

\makeatletter
\def\lst@visiblespace{\mbox{\kern.1em\vrule height.6ex}%
  \vbox{\hrule width1ex}%
  \hbox{\vrule height.6ex}%
}
\makeatother
\newcommand{\lstsetforpython}
{\lstset{language=Python,frame=single,frameround=tttt,framesep=3pt,backgroundcolor=\color{mygreen!20},rulecolor=\color{mygreen}}}

\usepackage{tikz} 
\usetikzlibrary{arrows.meta}
\usetikzlibrary{calc}
\usetikzlibrary{decorations.text}
\usetikzlibrary{positioning}
\usetikzlibrary{shapes, shapes.callouts}
\usetikzlibrary{trees}
\tikzset{
  >={Stealth}, 
  invisible/.style={opacity=0},
  visible on/.style={alt=#1{}{invisible}},
  alt/.code args={<#1>#2#3}{%
    \alt<#1>{\pgfkeysalso{#2}}{\pgfkeysalso{#3}} 
  },
}

\newcolumntype{C}[1]{>{\centering\arraybackslash}m{#1}}   
\newcolumntype{L}[1]{>{\raggedright\arraybackslash}m{#1}} 
\newcolumntype{R}[1]{>{\raggedleft\arraybackslash}m{#1}}  

\newcolumntype{+}{>{\global\let\currentrowstyle\relax}}
\newcolumntype{^}{>{\currentrowstyle}}
\newcommand{\rowstyle}[1]{\gdef\currentrowstyle{#1}#1\ignorespaces}

\makeatletter
\@ifundefined{htlatex}{%
  \usepackage{ctable}
  \setupctable{%
    botcap, 
    captionskip=2pt, 
    mincapwidth=2in, 
    framebg=1 1 1, 
    framefg=0 0 0, 
    framerule=0pt, 
    framesep=0pt,  
  }
}
\makeatother

\usepackage{colortbl}

\def\citea{\citet}

\newcommand{\parab}[1] {\bigskip \noindent \textbf{#1}~}
\newcommand{\param}[1] {\medskip \noindent \textbf{#1}~}
\newcommand{\paras}[1] {\smallskip \noindent \textbf{#1}~}

\newcommand{\setspacing}[1] {\linespread{#1}\selectfont}
\def\spacing {\setspacing}

\newlength{\parindentincr}
\newcommand{\addindent} {\addtolength{\parindent}{5mm}}
\newcommand{\indentby}[1] {
  \settowidth{\parindentincr}{#1}
  \addtolength{\parindent}{\parindentincr}}
\newcommand{\deindent} {\addtolength{\parindent}{-5mm}}
\newcommand{\deindentby}[1] {
  \settowidth{\parindentincr}{#1}
  \addtolength{\parindent}{-\parindentincr}}

\newcommand{\blank}[1] {\rule{#1}{0.5pt}}

\newcounter{zzlinei}
\newcommand{\blanklines}[1] {%
  \setcounter{zzlinei}{0}
  \whileboolexpr%
    {test{\ifnumcomp{\value{zzlinei}}{<}{#1}}}
    {\vspace{8mm}\hrulefill\par \addtocounter{zzlinei}{1}}
  }

\newcommand{\cl} {\centerline}
\newcommand{\command}[1] {\textcolor{Mahogany}{\texttt{#1}}}
\newcommand{\dash} {\blank}
\newcommand{\definedas} {\ensuremath{\triangleq\ }}
\newcommand{\dfslab} {\url{http://www.isical.ac.in/~dfslab}}
\newcommand{\emptybox} {\raisebox{3.3pt}{$\boxed{}$}}
\newcommand{\checkedbox} {\raisebox{3.3pt}{$\boxed{}$}\hspace{-7pt}$\checkmark$}
\newcommand{\greentick} {\includegraphics[scale=0.1]{green-tick}}
\providecommand{\half}{}
\renewcommand{\half} {\ensuremath{\frac{1}{2}}}
\renewcommand{\hl}[2][YellowOrange!50] {\colorbox{#1}{#2}}
\newcommand{\hlg}[1] {\colorbox{mygreen}{#1}}
\newcommand{\hlr}[1] {\colorbox{myred}{#1}}
\newcommand{\important}[1] {\centerline{\framebox{\textcolor{red}{#1}}}}
\newcommand{\intersection} {\cap}
\newcommand{\latex} {\LaTeX\ }
\newcommand{\lb} {\linebreak}
\renewcommand{\marks}[1] {\hspace*{\fill}{[#1]}}
\newcommand{\mc} {\multicolumn}
\newcommand{\mybox}[1] {\centerline{\framebox{#1}}}
\newcommand{\mycross} {\includegraphics[scale=0.1]{red-cross}}
\newcommand{\myellipsis}[3] {\newcount \secondindex
  \secondindex #2\relax
  \advance\secondindex +1\relax
  \ensuremath{{#1}_{#2},}
  \ensuremath{{#1}_{\the\secondindex},}
  \ensuremath{\ldots,}
  \ensuremath{{#1}_{#3}}%
}
\newcommand{\mysignature} {\includegraphics[scale=0.2]{signature}}
\newcommand{\mystrut}[1] {\rule[- #1 * \real{0.3}]{0pt}{#1}}
\newcommand{\otoprule} {\midrule[\heavyrulewidth]}
\providecommand{\overbar}{}
\renewcommand{\overbar} {\overline}
\newcommand{\partialinput}[2] {\ExecuteMetaData[#2]{#1}}
\NewCommandCopy{\pic}{\includegraphics}
\newcommand{\pipe} {~\ensuremath{\mid}~}
\newcommand{\pto} {\vfill\hspace*{\fill}\textsc{p.t.o.} \newpage}
\newcommand{\ra} {\ensuremath{\rightarrow}}
\newcommand{\Ra} {\ensuremath{\Rightarrow}}
\newcommand{\rcomment}[1] {\bigskip\color{blue}#1\color{black}}
\newcommand{\response} {\medskip\textbf{Response:~}}
\newcommand*{\rectgolla}[2][red]{
  \setlength{\fboxrule}{1pt}
  \fcolorbox{#1}{white}{#2}
} 
\newcommand{\sbcb} {\;\!} 
\renewcommand{\show} {\visible}
\renewcommand{\so} {\st}
\newcommand{\strikethrough} {\st}
\newcommand{\tb} {\textbackslash}
\newcommand{\torf} {\hfill \textsc{true / false}}
\newcommand{\tpgridon}{\TPShowGrid*{15}{12}} 
\newcommand{\ttilde} {\textasciitilde}
\newcommand{\undertilde}[1] {\mbox{\Large \raisebox{-5mm}{$\displaystyle\stackrel{#1}{\text{\textasciitilde}}$}}}
\newcommand{\union} {\cup}
\NewCommandCopy{\oldvec}{\vec}
\renewcommand{\vec}[1] {\oldvec{\vphantom{b}#1}}
\newcommand{\xor} {\ensuremath{\oplus}}

\newcommand{\anchorpoint}[1]{\tikz[overlay,remember picture] \node (#1) {};}

\newcommand*{\floatingnote}[5][1-]{
  \begin{tikzpicture}[overlay,remember picture]
    \pgftransformshift{\pgfpointanchor{current page}{center}}
    \path<#1> #2
    node[rectangle,rounded corners=4pt,fill=#4,opacity=.5]
    {\parbox{#3}{\raggedright #5}}
    ;
  \end{tikzpicture}
}

\newcommand*{\golla}[4][ellipse]{
  \begin{tikzpicture}[overlay,thick]
    \node () at #2 [shape=#1,draw=red,text height=#4] {~\hspace{#3}~} ;
  \end{tikzpicture}
}
 
\newcommand*{\namedgolla}[5][red]{
  \begin{tikzpicture}[overlay,remember picture,thick]
    \node (#5) at #2 [shape=ellipse,draw=#1,text height=#4] {~\hspace{#3}~} ;
  \end{tikzpicture}
}

\newcommand*{\popupnote}[4][1-]{
  \begin{tikzpicture}[overlay,remember picture]
    \pgftransformshift{\pgfpointanchor{current page}{center}}
    \path<#1> (#2) ++#3
    node[anchor=west,rectangle callout,rounded corners=4pt,fill=blue!50,
         opacity=.5,callout absolute pointer={(#2)}] {#4};
  \end{tikzpicture}
}

\newcommand*{\popupnotewithap}[5][1-]{
  \begin{tikzpicture}[overlay,remember picture]
    \pgftransformshift{\pgfpointanchor{current page}{center}}
    \path<#1> (#2) ++#3
    node[anchor=#5,rectangle callout,rounded corners=4pt,fill=blue!50,
         opacity=.5,callout absolute pointer={(#2)}] {#4};
  \end{tikzpicture}
}

\makeatletter
\newread\pin@file
\newcounter{pinlineno}
\newcommand\pin@accu{}
\newcommand\pin@ext{pintmp}
\newcommand*\dummypartialinput [3] {%
  \IfFileExists{#3}{%
    \openin\pin@file #3
    \setcounter{pinlineno}{1}
    \@whilenum\value{pinlineno}<#1 \do{%
      \read\pin@file to\pin@line
      \stepcounter{pinlineno}%
    }
    \addtocounter{pinlineno}{-1}
    \let\pin@accu\empty
    \begingroup
    \endlinechar\newlinechar
    \@whilenum\value{pinlineno}<#2 \do{%
      \readline\pin@file to\pin@line
      \edef\pin@accu{\pin@accu\pin@line}%
      \stepcounter{pinlineno}%
    }
    \closein\pin@file
    \expandafter\endgroup
    \scantokens\expandafter{\pin@accu}%
  }{%
    \errmessage{File `#3' doesn't exist!}%
  }%
}
\makeatother

\usepackage{adjustbox}
\let\Bbbk\relax
\usepackage{amsmath,amssymb}
\usepackage{multirow}
\usepackage{subcaption}

\definecolor{rm3}{HTML}{1F77B4}
\definecolor{bo1}{HTML}{FF7F0E}
\definecolor{kl}{HTML}{2CA02C}
\definecolor{ceqe}{HTML}{D62728}
\definecolor{hyde}{HTML}{9467BD}
\definecolor{hyderocchio}{HTML}{7F7F7F}

\title{Towards Explaining Query Expansion Performance in Information Retrieval}
\date{}

\begin{document}

\author{Sourav Saha}
\affiliation{%
	\institution{Indian Statistical Institute}
	\city{Kolkata}
	\country{India}
}
\email{sourav.saha\_r@isical.ac.in}

\author{Aditya Dutta}
\affiliation{%
	\institution{IIT Bombay}
	\city{Bombay}
	\country{India}
}
\email{aditya.dutta@iitb.ac.in}

\author{Soumajit Pramanik}
\affiliation{%
	\institution{IIT Bhilai}
	\city{Chhattisgarh}
	\country{India}
}
\email{soumajit.pramanik@gmail.com}

\author{Mandar Mitra}
\affiliation{%
	\institution{Indian Statistical Institute}
	\city{Kolkata}
	\country{India}
}
\email{mandar@isical.ac.in}

\begin{abstract}
Query Expansion (QE) techniques have long been widely used in Information Retrieval (IR) to address the \textit{vocabulary mismatch problem}. They remain relevant in modern retrieval systems, including those based on large language models (LLMs).
However, no single QE method consistently outperforms others across all queries.
This work seeks to explain the variation in QE performance through two complementary perspectives. The first is the concept of an \textit{Ideal Expanded Query (IEQ)}---a hypothetical query that maximizes retrieval effectiveness with a downstream BM25 retrieval model. The second is a \textit{separability perspective}, which quantifies how distinctly relevant and non-relevant documents are scored for a given expanded query using Cohen's (d). We develop a separability measure and practical formulations to approximate the IEQ and investigate how these factors relate to retrieval effectiveness. Extensive experiments on the TREC Robust collection, TREC DL 2019--2022 passage collections, and TREC DL 2019--2020 document collections reveal several interesting patterns. In particular, we find that expanded queries that are closer to the ideal expanded query tend to achieve higher retrieval effectiveness. We further show that the separability of relevant and non-relevant documents provides a complementary perspective for understanding QE performance.

\end{abstract}



\maketitle


\section{Introduction}
\label{introduction}
The primary motivation behind Query Expansion (QE) techniques is to address
one of the fundamental problems in Information Retrieval (IR), namely the
vocabulary mismatch problem, where searchers and authors of relevant
content use different terms to refer to the same concept. The primary
objective of QE is to enrich query representations by adding related terms
in order to reduce the chances of vocabulary mismatch, and thereby improve
retrieval effectiveness. Numerous successful automatic QE techniques have
been proposed over the years~\cite{qe-survey-csur}. Research on automatic
QE remains popular even after the emergence of neural ranking methods,
semantic search, and the use of large language models (LLMs) in
IR~\cite{qe-survey-tois}.

Because QE methods generally yield significant improvements on average
(across a set of queries), they are widely applied in IR. It is well-known,
however, that in unsupervised settings, e.g., pseudo relevance feedback
(PRF), these approaches exhibit highly variable performance across queries:
for a non-trivial number of queries, automatic expansion results in
degraded performance compared to the original query. Further, two QE
methods that are about equally effective on average, may have very
different performance characteristics across individual queries. This
variability raises an important question: what makes one expanded query
more effective than another? Prior work has largely focused on improving QE
effectiveness, while relatively little attention has been paid to
understanding the variations across QE methods and queries.

In this study, our goal is 
to propose a \emph{post hoc} approach for analysing QE methods in order to
better understand {why} QE methods behave differently for different
queries. More precisely, we aim to uncover interpretable properties that
can systematically account for variations in the effectiveness of a single
method across queries, and of different QE methods on the same query. Our
explanations are intended for IR researchers and practitioners, rather than
for end users of IR systems. The proposed approach is specifically targeted
at a researcher~/ developer who is trying to understand in intuitive terms
why a QE method fails (or does well) on a particular query.

Traditional QE methods typically involve two components: (i)~term selection
--- determining which terms should be added to the query, and (ii)~term
weighting --- assigning numeric importance scores to the selected terms
that are then used by a retrieval model to compute query-document
similarity scores. Thus, an IR researcher may use the following hypothesis
as a basis for analysing QE methods.
\begin{quotation}
  A QE method that performs well does so because it (i)~selects good terms,
  and (ii)~assigns appropriate weights to the terms. Conversely, a proposed
  QE algorithm that turns out to be ineffective is likely to have failed
  because (i)~it fails to identify useful expansion terms for inclusion,
  and/or (ii)~it assigns inappropriate weights to expansion terms.
\end{quotation}
We conceptualise \emph{ideal expanded queries} as a means to jointly
formalise the notion of ``good'' expansion terms, and ``good'' expansion
weights. Given an information need, an ideal expanded query (IEQ) is a
query vector that would yield (for a given retrieval model) perfect, or
near perfect, Average Precision (AP). We then study how closely different
expanded queries resemble this ideal, and how such similarity correlates
with their observed retrieval performance.

Our notion of an ideal query is rooted in traditional scoring functions
that are in turn based on keyword matching. At first glance, this may
appear to be a weakness, since these models have mostly been superseded by
neural retrieval models that use more sophisticated scoring functions that
are better at capturing context-aware semantic matches. We emphasise,
however, that this choice is valid: it is standard (indeed, recommended)
practice to use traditional methods and models (that are generally better
understood) to explain more complex ones. Thus, our approach is entirely
consistent with an approach widely used in the explainability literature.

In addition to the IEQ perspective, we also consider another related
perspective: we view retrieval as a discrimination problem between relevant
and non-relevant documents. We hypothesise that an expanded query (EQ)
performs well when it produces a strong statistical separation between
their BM25 scores. While this relationship is intuitively obvious, we
empirically investigate the extent to which separability is associated with
retrieval effectiveness.

To summarise, this study examines the following research questions.
\begin{itemize}
\item[\textbf{RQ1 -}] For a given query $Q$, consider different expanded
  versions 
  $\mathbf{Q}^{(1)}_{\mathrm{exp}}, \mathbf{Q}^{(2)}_{\mathrm{exp}},
  \ldots$ (such as those generated by different QE methods), and the
  \emph{ideal expanded query}. For these EQs, how strongly does similarity
  to the IEQ correlate with retrieval effectiveness? On a related note, if
  an EQ is moved toward or away from the ideal query, how does its
  retrieval effectiveness vary?

\item[\textbf{RQ2 -}] How strongly does an EQ's ability to separate the
  scores of relevant and non-relevant documents correlate with its
  effectiveness? 

\item[\textbf{RQ3 -}] Can similarity to the ideal query account for the
  variability of a QE technique across queries? Specifically, can it
  provide additional information when automatic QE techniques fail to
  improve retrieval effectiveness?
\end{itemize}
To the best of our knowledge, this work is the first to consider the
performance of expanded queries from this perspective. We conducted
extensive experiments on the Robust, TREC DL 2019, and TREC DL 2020
document collections, as well as the TREC DL 2019, 2020, 2021 and 2022
passage collections. The results suggest that all the questions above have
positive answers.

We emphasise that our framework provides \emph{actionable}
  explanations in the following sense. When a QE method $\mathcal{M}$ fails
  for a particular query $Q$, our framework helps a researcher to identify
  whether $Q$ is generally hard for automatic QE approaches. Next, it helps
  to identify specific expansion terms that $\mathcal{M}$ fails to select,
  or assigns inappropriate weights to. This provides a concrete starting
  point for exploring how $\mathcal{M}$'s shortcomings may be addressed.

The remainder of this paper is organized as follows. Section~\ref{rel-work}
reviews related work. Section~\ref{proposed-framwork} presents our proposed
framework for analyzing and explaining QE methods.
Section~\ref{experimental-setup} describes the experimental setup,
Section~\ref{results} presents our experimental results, and
Section~\ref{sec:discussion} provides an analysis of our experimental
findings. Finally, Section~\ref{conclusion-future-work} discusses
limitations and directions for future work. To aid reproducibility, we release our code and results to the community at \url{https://github.com/souravsaha/py-qe-terrier}.


\section{Related work}
\label{rel-work}
Existing QE research primarily focuses on developing more effective
expansion techniques rather than analyzing the mechanisms underlying their
success. To the best of our knowledge, no prior work has systematically
investigated the effectiveness of different QE methods on individual
queries. Below, we provide a brief review of research on query expansion,
explainable IR, and the interplay between these two areas.

\subsection{Query expansion}
\label{rel-work-qe}
Query expansion (QE) has been extensively studied by the IR community, in
the context of applications spanning from traditional keyword-based
document search to modern question-answering systems. As mentioned in the
Introduction, traditional statistical IR models addressed the vocabulary
mismatch problem by
selecting additional terms to augment the original query, and assigning
appropriate weights to the selected terms. Consequently, a large body of
research focused on improving QE effectiveness through better term
selection and weighting strategies. Representative approaches include
Rocchio~\cite{Rocchio1971} and its unsupervised variant, \emph{blind} or
\emph{pseudo} relevance feedback (PRF), the relevance-based language models
RM1~\cite{rm1} and RM3~\cite{rm3}, and Bose-Einstein weighting
(Bo1~\cite{bo1_kl}). A comprehensive survey of traditional QE techniques is
provided in~\cite{qe-survey-csur}.

With the advent of deep learning, various neural representations have been
incorporated into QE methods. Early neural approaches leveraged static word
embeddings, such as Word2Vec~\cite{word2vec}, for selecting expansion
terms~\cite{roy2016using,diaz-etal-2016-query}. Subsequently, researchers
explored contextualized representations derived from transformer-based
models. For example, CEQE~\cite{ceq-naseri-ecir-21} integrates BERT-based
contextual embeddings into the RM3 framework to improve expansion term
selection and enrich query representations, resulting in substantial
improvements in retrieval effectiveness.

LLMs have also emerged as powerful sources of external knowledge
for query expansion. Several
approaches~\cite{mugi-emnlp-2024-exploring-best, wang-etal-2023-query2doc,
  feedback_models_for_hyde, gao-etal-2023-precise} generate synthetic
pseudo-relevant documents by prompting LLMs and leverage them for query
reformulation. Researchers have also explored alternative prompting
strategies~\cite{jagerman2023queryexpansionpromptinglarge} and different
forms of expansion content, such as entities and
summaries~\cite{dhole-ecir-24}. More recently,
ThinkQE~\cite{lei-etal-2025-thinkqe} proposed an iterative refinement
framework that progressively improves expansion terms through multi-step
reasoning. 

Despite these advances in LLM based expansion generation, the weighting of
expansion terms often remains heuristic. Common strategies include simple
concatenation of generated content with the original query or repetition of
query terms before concatenation~\cite{mugi-emnlp-2024-exploring-best}. In
contrast, \citet{feedback_models_for_hyde} demonstrated that classical
weighting schemes, such as Rocchio and RM3, remain highly effective for
weighting LLM-based expanded queries. A recent survey~\cite{qe-survey-tois}
provides a comprehensive overview of modern QE techniques and their applications across domains such as web search, medical IR, conversational search, and code search.

\subsection{Explainable IR}
\label{rel-work-exir}
Modern IR systems have greatly advanced the state of the art in retrieval
effectiveness, but they rely increasingly on complex neural architectures,
often making them black-boxes. Consequently, there has been growing
interest in understanding and explaining their behavior. Researchers have
proposed both post-hoc explanation techniques~\cite{lirme, exs} and
retrieval models that are interpretable by
design~\cite{leonhard-explainable-by-design}. In this context, two
surveys~\cite{explain-ir-avishek-anand-survey, explainable-survey-acm-csur}
and tutorials~\cite{sigir-23-tutorial, exir_ecir_tutorial} provide comprehensive overviews of
explainable IR and establish key terminology in this area. A toolkit that
implements common approaches to explainable IR, as well as methods for
evaluating them, is described in~\cite{ir_explain_sigir}. 

One important dimension of explainability concerns the intended audience of
the explanation. Explanations may be designed for end users to improve
transparency and trust in the retrieval system. Alternatively, explanations
may target IR system developers, helping them understand the sources of
effectiveness gains or performance differences among retrieval models. For
example, \citet{i-rex} focus on developer-oriented explanations, while the
analysis in~\cite{smucker2005investigation} helps IR practitioners
understand that the language modeling based scoring method tends to favor
shorter documents when Jelinek-Mercer
smoothing~\cite{zhai-lafferty-smoothing-tois-2004} is used instead of
Dirichlet smoothing~\cite{zhai-lafferty-smoothing-tois-2004}. Such
diagnostic analyses are particularly important for modern neural retrieval
systems, whose behavior is often difficult to interpret.
Our work in this study falls within the second category, in that our
explanations are targeted at IR researchers. 
We believe our approach provides researchers with tools that help diagnose,
compare, and improve both traditional and modern QE techniques.

\subsection{Query Expansion applied to Explainable IR and vice versa}
\label{rel-work-exir}
Despite a growing body of literature on Explainable IR, relatively little
attention has been paid to understanding when and why QE methods improve
retrieval effectiveness. The Axiomatic
Framework~\cite{fang-sigir-04,semantic-term-matching-fang,fang-tao-zhai-tois-11}
may be regarded as one of the earliest approaches to Explainable IR in
general --- it formalises intuitive ideas about when IR systems
(term-weighting models, in particular) work well, and helps to explain the
drawbacks of certain term-weighting models.
\citet{clinchant2013theoretical} extended this framework to pseudo
relevance feedback-based QE techniques. Their work helps to explain why
certain QE models are less effective than others \emph{on the whole}; such
explanations are classifed as \emph{global} explanations. In contrast, our
objective is to provide \emph{local} explanations that help
understand why a QE method works well or poorly for a specific query.

Even though research in the area of Explainable IR has not focused much on
analysing QE models and methods, QE techniques have been used quite often
as a component within explanatory frameworks. One common approach used by
the Explainability community involves interpreting complex models as a
conglomerate of simple models, each of which behaves like a local
approximation of the complex model within a restricted region~\cite{lime}.
This idea has been applied to neural retrieval models as
follows~\cite{lirme,lijun-ecir-23}. Neural rankers are believed to perform
well because they identify semantic matches (rather than simple lexical
overlap) between queries and documents. Given a query $Q$, if a
classical retrieval model (like BM25), operating on
$\mathbf{Q}_{\mathrm{exp}}$, an expanded version of $Q$, produces
a ranking that is similar to that produced by a neural ranker $M$ from
$Q$, we may hypothesise that $M$ performs semantic matching by
implicitly exploiting these specific expansion terms and concepts during
document ranking. In such approaches, the quality of the explanation is
measured by the rank correlation between the rankings produced by the two
models, with higher correlation indicating a more faithful explanation.

\section{Proposed framework}
\label{proposed-framwork}
In this section, we present our framework for explaining the performance of
QE methods. The framework consists of three main components:
(1)~constructing an ideal expanded query, \(\mathbf{Q}_{\mathrm{IEQ}}\)
(Section~\ref{construction-ieq}), (2)~measuring the similarity between
\(\mathbf{Q}_{\mathrm{IEQ}}\) and an expanded query
\(\mathbf{Q}_{\mathrm{exp}}\)
(Section~\ref{pairwise-listwise-corr-define}), and (3)~quantifying how well
\(\mathbf{Q}_{\mathrm{IEQ}}\) or an expanded query
\(\mathbf{Q}_{\mathrm{exp}}\) distinguishes between top-ranked relevant and
non-relevant documents (Section~\ref{separability-formulation}).

Formally, let $Q$ be a user query, $C$ be the document collection, and $V$
= $\{\myellipsis{t}{1}{|V|}\}$ be the vocabulary of $C$. Our first
component retrospectively constructs \(\mathbf{Q}_{\mathrm{IEQ}}\) as a
query representation constructed from terms in $V$. Our objective is that
\(\mathbf{Q}_{\mathrm{IEQ}}\) should achieve near-optimal retrieval
effectiveness, i.e., AP close to 1.0. As explained in Section~1, our
expectation is that \(\mathbf{Q}_{\mathrm{IEQ}}\) would contain all and
only the ``good'' terms with ``optimal'' weights.

The second component addresses \textbf{RQ1} by looking at the correlation
between the proximity of \(\mathbf{Q}_{\mathrm{exp}}\) to
\(\mathbf{Q}_{\mathrm{IEQ}}\) and its retrieval effectiveness. The
underlying intuition is that if \(\mathbf{Q}_{\mathrm{exp}}\) contains a
larger proportion of good terms with appropriate weights, it will be more
similar to \(\mathbf{Q}_{\mathrm{IEQ}}\), and should yield better retrieval
effectiveness.

\textbf{RQ2} is addressed by the third component, which evaluates an EQ
based on how well it separates top-ranked relevant and non-relevant
documents. Our intuition is that an effective EQ should assign more
distinct scores to relevant and non-relevant documents than an ineffective
EQ, i.e., if $\mathbf{QE}_1$ achieves better retrieval effectiveness than
$\mathbf{QE}_2$, then $\mathbf{QE}_1$ is expected to induce greater
separation between the scores of relevant and non-relevant documents. To
quantify this effect, we employ Cohen's (d) as a measure of separability;
Section~\ref{separability-formulation} describes this in more detail.


\subsection{Construction of \(\mathbf{Q}_{\mathrm{IEQ}}\)}
\label{construction-ieq}
An ideal expanded query (\(\mathbf{Q}_{\mathrm{IEQ}}\)) is a conceptual
construct. It can be approximated empirically using complete relevance
information for a given query. We explore two approaches that leverage
available relevance judgments to construct approximate instantiations of
\(\mathbf{Q}_{\mathrm{IEQ}}\): (1) a Dynamic Feedback
Optimization~\cite{dfo-buckley-sigir-95} (DFO) method that iteratively
adjusts term weights to directly maximize average precision (AP), and (2) a Least-Squares (LS)
surrogate objective that encourages high retrieval scores for relevant
documents. These two instantiations are denoted as
$\mathbf{Q}_{\mathrm{IEQ}}^{\mathrm{DFO}}$ and
$\mathbf{Q}_{\mathrm{IEQ}}^{\mathrm{LS}}$ respectively.
To the best of our knowledge, constructing ideal queries using a
least-squares surrogate objective represents a novel perspective that has
not been explored in prior work.

\subsubsection{Dynamic Feedback Optimization (DFO)~\cite{dfo-buckley-sigir-95}}
\label{dfo-section}
DFO was originally proposed as a method for tuning the term weights of an
EQ using training data (relevance judgments on a training collection). In
this work, we use DFO as a component of a \emph{post hoc} explanation
method. Thus, in order to construct
$\mathbf{Q}_{\mathrm{IEQ}}^{\mathrm{DFO}}$, we directly use the full
relevance information for the target collection itself, rather than a
training corpus, to tune the weights of the expansion terms.

We begin with \(\mathbf{Q}_\mathrm{rocchio}\), a query vector constructed
by applying Rocchio Relevance Feedback. We then iteratively refine the
expansion term weights with the objective of maximizing AP as follows.

\begin{enumerate}[I.]
\item Truncate \(\mathbf{Q}_\mathrm{rocchio}\) to the top $r$ positively
  weighted terms.
\item Sort the terms according to their final weights \(w_t\).
\item For each magnitude \(m \in \{4.0, 2.0, 1.0, 0.5, 0.25\}\),
  \begin{itemize}
  \item For each term $t$ in \(\mathbf{Q}_\mathrm{rocchio}\), considered in
    decreasing order by weight (\(w_t\)):
    \begin{itemize}
    \item Update \(w_t \leftarrow (1+m)w_t\).
    \item Retain the updated weight only if it improves AP; otherwise,
      revert to the previous value.
    \end{itemize}
  \end{itemize}
\end{enumerate}
We retain only the top $r$ terms with the highest weights in Step I because
DFO involves multiple retrieval iterations. Choosing a moderate value for
$r$ (e.g., 200) ensures that the method completes within a reasonable
amount of time. The magnitude vector in Step III is adopted directly
from~\cite{dfo-buckley-sigir-95}.

\subsubsection{Least-Squares Surrogate}
\label{ls-surrogate}
More generally, we could regard AP as a function $\mathcal{F}$ of term
weights. Constructing \(\mathbf{Q}_{\mathrm{IEQ}}\) is equivalent to
searching for points in the space of possible term weights where
$\mathcal{F}$ is maximum. From this perspective, DFO represents one
particular search strategy. Directly optimising AP (using gradient descent,
for example) is not practical, because $\mathcal{F}$ does not have a
convenient, differentiable form. We therefore seek a differentiable
surrogate objective that encourages separation between relevant and
non-relevant documents. Let document $D_j$ be
represented as a vector:
\begin{equation}
  \label{eq:docvec}
  \mathbf{D}_j = (w_1^{(j)}, w_2^{(j)}, \ldots, w_{|V|}^{{(j)}})^{\top},
\end{equation}
where $w_i^{(j)}$ denotes the BM25 weight of the term $t_i$ in the document
$D_j$. Let \(\mathbf{X}\in\mathbb{R}^{n\times |V|}\) denote the term-document matrix
for the collection $C$ ($n = |C|$, and the $j$-th row of $\mathbf{X}$ corresponds to
$\mathbf{D}_j^\top$).

For a query vector \(\mathbf{Q} \in \mathbb{R}^{|V|}\), the product
\(\mathbf{X}\mathbf{Q}\) is the vector of BM25 scores for all documents in
$C$.\footnote{Both \(\mathbf{Q}\) and \(\mathbf{D}\) are represented in a
  shared term space. This idea is also central to learned sparse retrieval
  methods such as SPLADE~\cite{splade, splade_v2}. However, unlike these
  approaches, our goal is not to learn sparse query and document
  representations for retrieval, but to construct and analyze ideal query
  representations using full relevance information.} We define a target
vector \(\mathbf{Y}\in\mathbb{R}^n\) using relevance labels as follows:
\[
  Y_j =
  \begin{cases}
    \alpha & \text{if document } D_j \text{ is relevant},\\[4pt]
    0 & \text{if document } D_j \text{ is non-relevant},
  \end{cases}
\]
where \(\alpha>0\) is a constant. We then seek to obtain
$\mathbf{Q}_{\mathrm{IEQ}}^{\mathrm{LS}}$ as the solution to the following
optimization problem:
\begin{equation}
  \label{eq:ls-optimization}
  \mathbf{Q}_{\mathrm{IEQ}}^{\mathrm{LS}} = \arg\min_{\mathbf{Q}} \; \big\|\mathbf{X}\mathbf{Q} - \mathbf{Y}\big\|_2^2 + \lambda \|\mathbf{Q}\|_2^2,
\end{equation}
where $\lambda$ is a regularization parameter. The solution to the
above~\cite[Chapter 3]{hastie2009elements} is given by
\begin{equation}
  \label{eq:ls-solution}
  \mathbf{Q}_{\mathrm{IEQ}}^{\mathrm{LS}} = (\mathbf{X}^\top \mathbf{X} +
  \lambda I)^{-1} \mathbf{X}^\top \mathbf{Y}.
\end{equation}
($I$ is the identity matrix). An important property of this formulation is
that the resulting AP is invariant to the specific positive value chosen
for \(\alpha\) (see Appendix~\ref{sec:invariance} for details).
  For typical text collections, however, $|V|$ is too large for
  the calculation of $(\mathbf{X}^\top \mathbf{X} + \lambda I)^{-1}$ to be
  practicable. Thus, in applications, the optimisation problem in
  Equation~\ref{eq:ls-optimization} is solved using gradient descent methods.

\subsection{Relation between an EQ's similarity to
  \(\mathbf{Q}_{\mathrm{IEQ}}\) and its retrieval effectiveness}
\label{pairwise-listwise-corr-define}
Having constructed approximate IEQs
$\mathbf{Q}_{\mathrm{IEQ}}^{\mathrm{DFO}}$ and
$\mathbf{Q}_{\mathrm{IEQ}}^{\mathrm{LS}}$, we next attempt to answer
\textbf{RQ1}. Given an EQ \(\mathbf{Q}_{\mathrm{exp}}\), we measure how
closely it aligns with \(\mathbf{Q}_{\mathrm{IEQ}}\), and analyze the
correlation between this similarity and the retrieval effectiveness (AP
score) achieved by \(\mathbf{Q}_{\mathrm{exp}}\). We use the standard
cosine similarity formula to measure the similarity between
\(\mathbf{Q}_{\mathrm{exp}}\) and \(\mathbf{Q}_{\mathrm{IEQ}}\):
\begin{equation}
  \mathrm{Sim}(\mathbf{Q}_{\mathrm{exp}}, \mathbf{Q}_{\mathrm{IEQ}})
  =
  \frac{\mathbf{Q}_{\mathrm{exp}}^\top \mathbf{Q}_{\mathrm{IEQ}}}
  {\|\mathbf{Q}_{\mathrm{exp}}\|_2 \|\mathbf{Q}_{\mathrm{IEQ}}\|_2}.
\end{equation}

For a particular user query $Q$, let
\[
  \{\mathbf{Q}^{(1)}_{\mathrm{exp}}, \mathbf{Q}^{(2)}_{\mathrm{exp}}, \dots,
  \mathbf{Q}^{(k)}_{\mathrm{exp}} \}
\]
denote a set of $k$ EQs generated for $Q$ by different QE algorithms and/or
parameter settings. For each variant \(\mathbf{Q}^{(i)}_{\mathrm{exp}}\),
we compute both its similarity score Sim$(\mathbf{Q}^{(i)}_{\mathrm{exp}},
\mathbf{Q}_{\mathrm{IEQ}})$
and its retrieval
effectiveness $AP(\mathbf{Q}^{(i)}_{\mathrm{exp}})$. We then compute
the correlation between the $k$ similarity scores and corresponding AP
values:
\begin{equation}
  \rho_Q = \mathrm{corr} \big( \{
  \mathrm{Sim}(\mathbf{Q}^{(i)}_{\mathrm{exp}}, \mathbf{Q}_{\mathrm{IEQ}})
  \}_{i=1}^k, ~~ \{ AP(\mathbf{Q}^{(i)}_{\mathrm{exp}})\}_{i=1}^k
  \big)
\end{equation}
where \(\mathrm{corr}\) denotes either Pearson's $\rho$ or Spearman's rank
correlation. This yields one correlation value \(\rho_Q\) per query.
Finally, we report the mean correlation across all queries:
\begin{equation}
  \label{avg-corr-sim-to-different-qe}
  \bar{\rho}
  =
  \frac{1}{|\mathcal{Q}|}
  \sum_{Q \in \mathcal{Q}} \rho_Q.
\end{equation}

\subsection{How well can an EQ separate relevant and non-relevant documents?}
\label{separability-formulation}
Next, we develop our second perspective on understanding the performance
differences among EQs generated by different QE methods. We quantify how
well an EQ separates relevant documents from non-relevant ones. 
Section~\ref{separability-with-cohens-distribution}, we 
quantify the difference in the distribution of BM25 scores (with respect to a given
\(\mathbf{Q}_{\mathrm{exp}}\)) across the relevant and (judged)
non-relevant document sets for this query. 

\subsubsection{Separability of BM25 scores}
\label{separability-with-cohens-distribution}
Following the notation introduced in Section~\ref{ls-surrogate}, we use
\(\mathbf{X}^{(Q)}_\mathrm{Rel}\) and \(\mathbf{X}^{(Q)}_\mathrm{NRel}\) to denote the
term-document matrices for the set of all \emph{judged} relevant and
non-relevant documents, respectively, for a user query $Q$. For a given
\(\mathbf{Q}_{\mathrm{exp}}\) corresponding to $Q$, the BM25 scores of
these two sets of documents are given by the vectors
\(\mathbf{S}_\mathrm{Rel} = \mathbf{X}^{(Q)}_\mathrm{Rel}\mathbf{Q}_\mathrm{exp}\) and
\(\mathbf{S}_\mathrm{NRel} = \mathbf{X}^{(Q)}_\mathrm{NRel}\mathbf{Q}_\mathrm{exp}\)
respectively.
Intuition suggests that an effective EQ should assign substantially
higher scores to relevant documents than to non-relevant documents.
Conversely, an EQ that achieves a greater separation between the score
distributions \(\mathbf{S}_{\mathrm{Rel}}\) and
\(\mathbf{S}_{\mathrm{NRel}}\) is expected to be more similar to
\(\mathbf{Q}_{\mathrm{IEQ}}\).

A simple measure of the separation between \(\mathbf{S}_{\mathrm{Rel}}\)
and \(\mathbf{S}_{\mathrm{NRel}}\) would be $\mu_\mathrm{Rel} -
\mu_\mathrm{NRel},$ where $\mu_\mathrm{Rel}$ and $\mu_\mathrm{NRel}$
denote the means of $\mathbf{S}_\mathrm{Rel}$ and
$\mathbf{S}_\mathrm{NRel}$, respectively. \footnote{This is precisely the
  starting point of Rocchio's derivation of his relevance feedback
  algorithm.} Of course, this difference has to be viewed in the context of
the variability of the scores within \(\mathbf{S}_{\mathrm{Rel}}\) and
\(\mathbf{S}_{\mathrm{NRel}}\). This naturally leads to our choice of
\textit{Cohen's} \(d\)~\cite{cohen1988statistical} as a measure of the separation between the distribution of scores in $\mathbf{S}_\mathrm{Rel}$ and
$\mathbf{S}_\mathrm{NRel}$. Thus, we define
$\mathrm{Separability}(\mathbf{Q}_\mathrm{exp})$ as 
\begin{equation}
  \label{qe-score-separability-final}
  \mathrm{Separability}(\mathbf{Q}_\mathrm{exp}) = \frac{\mu_\mathrm{Rel} - \mu_\mathrm{NRel}}{\sqrt{\frac{(n_\mathrm{Rel}-1)\sigma_\mathrm{Rel}^2+(n_\mathrm{NRel}-1)\sigma_\mathrm{NRel}^2}{n_\mathrm{Rel}+n_\mathrm{NRel}-2}}}
\end{equation}
where $n_\mathrm{Rel}$ and $n_\mathrm{NRel}$ denote the numbers of relevant
and non-relevant documents, respectively, and $\sigma_\mathrm{Rel}^2$ and
$\sigma_\mathrm{NRel}^2$ denote the variances of their BM25 scores.

\subsubsection{Comparing the separability score of a Query Expansion to its similarity to \(\mathbf{Q}_{\mathrm{IEQ}}\)}
\label{separability-correlation}
As in Section~\ref{pairwise-listwise-corr-define}, we consider a set of $k$
EQs generated for a user query $Q$ using different QE algorithms and/or
parameter settings, and analyze the correlation between
\(\mathrm{Separability}(\mathbf{Q}_\mathrm{exp})\) measured using
Equation~\ref{qe-score-separability-final} 
and
AP across these $k$ variants. Finally, we report the average of these correlations
across all queries.



\section{Experimental setup}
\label{experimental-setup}
This section describes the experimental setup, including the query
expansion algorithms, datasets, and evaluation settings used in our
experiments.

\subsection{Datasets}
\label{datasets}
We use the following adhoc IR test collections: TREC Robust, MS MARCO v1
passage and document collections evaluated using the TREC Deep Learning
2019 and 2020 queries, and the MS MARCO v2 passage collection evaluated
using the TREC Deep Learning 2021 and 2022 queries. For brevity, we merge
the topics and relevance judgments from the TREC DL 2019 and 2020
evaluations and refer to the resulting benchmarks as DL19-20 Passage and
DL19-20 Document for the MS MARCO v1 passage and document collections,
respectively. Similarly, we combine the TREC DL 2021 and 2022 evaluations
for the MS MARCO v2 passage collection and refer to the resulting benchmark
as DL21-22 Passage. Dataset statistics are summarised in
Table~\ref{tab:datastats}.

\begin{table}[t]
  \centering
  \caption{\small Datasets used in our experiments. The genre column denotes the collection type. `avg.$|Q|$' and
    `avg.\#rel' denote average query length and average number of relevant
    documents, respectively. The `\#topics' column reports the total number
    of topics, while the values in parentheses indicate the number of
    topics in the individual query sets for the corresponding years.}

  \begin{adjustbox}{width=0.9\columnwidth}
    \begin{tabular}{l c c c c c} \toprule
      Collection & genre & \#docs & \#topics & avg.$|Q|$ & avg.\#rel \\\midrule

      \textbf{TREC Robust} & news & 528,155  & 249 & 2.62 & 68.36 \\

      \textbf{DL19-20 Passage} & web pages &   8,841,823  & 97 (43 + 54) & 5.76 & 42.96 \\
      \textbf{DL19-20 Document} & web pages & 3,213,835 &  88 (43 + 45) & 5.92 & 95.05 \\

      \textbf{DL21-22 Passage} & web pages &  138,364,198  & 129 (53 + 76) & 7.43 & 396.64  \\
      \bottomrule
    \end{tabular}
    \label{tab:datastats}
  \end{adjustbox}
\end{table}

\subsection{QE approaches explored}
\label{qe-approaches-explored}
We evaluate five different QE methods, covering statistical, neural and
LLM-based approaches. Traditional QE techniques include (i)~RM3~\cite{rm3},
(ii)~KL-divergence-based QE~\cite{bo1_kl}, and
(iii)~Bo1~\cite{bo1_kl}. To complement these classical methods, we also
include (iv)~CEQE~\cite{ceq-naseri-ecir-21}, a BERT-based contextualized
query expansion technique, and (v)~the feedback model for
HyDE~\cite{feedback_models_for_hyde}, an LLM-based method. For HyDE, we
consider two term-weighting variants: (a)~the Rocchio-based weighting
scheme, which was reported to be the most effective by \citet{feedback_models_for_hyde}
(HyDE-Rocchio), and (b)~a simple averaging scheme that combines the query
vector and feedback-document vector (HyDE).

\parab{Parameter settings.} Most QE methods have two important parameters:
$t_d$, the number of pseudo-relevant documents used for feedback, and $t_n$ the
number of expansion terms added to the original query. For all methods, we
vary $t_d$ over $\{10, 20, 30, 40\}$, and the number of expansion terms $t_n \in \{15, 25,
35, 45, 55\}$, in each case selecting the top-$t_n$ ranked expansion terms for inclusion in the query. This allows us to
generate a diverse set of expanded queries that differ both in the source
documents used for the expansion and in the selected expansion terms. Other
method-specific parameter settings are summarised below.

For RM3 and CEQE the mixing parameter $\mu$ is set to 0.6. 
We use the max-pooling variant of CEQE in our experiments. For further details, see~\citet{ceq-naseri-ecir-21}.
Following~\cite{feedback_models_for_hyde}, we generate \textbf{eight}
synthetic documents, each containing up to \textbf{512} tokens, using the
\textbf{Qwen-2.5-7B-Instruct}~\cite{yang2024qwen2} model. For HyDE-Rocchio, we set $\alpha = 1$ and $\beta = 0.75$, with $\gamma$ fixed at 0.

The above parameter settings for RM3, CEQE, KL-divergence and Bo1 result in
20 EQs (four options for $t_d$ and five possible values for $t_n$) for each
method, for a total of 80 EQ variants. For the two HyDE-based approaches,
the number of synthetic feedback documents is fixed, and only the number of
expansion terms is varied, resulting in 10 additional EQs. In the
experiments and analysis discussed in subsequent sections, we consider
these 90 EQ variants for each benchmark query.

\subsection{Other experimental details}
\label{exp-settings}
\subsubsection{Retrieval setup}
\label{retrieval-setup}
All indexing and retrieval experiments are conducted using
PyTerrier~\cite{pyterrier2020ictir}. We use BM25 as the downstream
retrieval model with its default parameter settings, \(k_1 = 1.2\) and \(b
= 0.75\). The Porter stemmer is used during indexing; expansion terms
generated with neural models are also stemmed before the final retrieval.

\subsubsection{Ideal Expanded Query Construction with DFO}
\label{exp-settings-ieq-dfo}
To construct the initial Rocchio query vector
($\mathbf{Q}_{\mathrm{rocchio}}$), we use $\alpha = 2.0, \beta = 64.0,
\gamma = 64.0.$ (multiplicative factors for the weights of terms from the
original query, positive feedback documents, and negative feedback
documents, respectively). Because each iteration of DFO requires full
retrieval followed by evaluation for the query, it is computationally a
very expensive process. Therefore, we include only the best 200 terms in
$\mathbf{Q}_{\mathrm{rocchio}}$, and terminate the iterative search when
the AP for the current query reaches or crosses 0.9.

\subsubsection{Ideal Query Construction with LS}
\label{exp-settings-ieq-ls}
\(\mathbf{Q}_{\mathrm{IEQ}}^{\mathrm{LS}}\) is obtained as the solution to
the optimisation problem stated in Equation~\eqref{eq:ls-optimization}, but
instead of using the full term-document matrix $\mathbf{X}$, we consider
$\mathbf{X}_{\mathrm{judged}}$, which consists of only the explicitly judged
documents. We use gradient descent for 5000
epochs, and employ a cosine annealing learning rate schedule with a maximum
learning rate of 0.1, decaying to a minimum value of 0. We try two values
for the regularization parameter: $\lambda=0.1$ and $\lambda=1$ (see Appendix~\ref{ablation-studies} for details).

The dimensionality of \(\mathbf{Q}_{\mathrm{IEQ}}^{\mathrm{LS}}\)
corresponds to the vocabulary size of the judged documents, resulting in a
dense query representation containing both positive and negative term
weights. To evaluate the retrieval effectiveness of
$\mathbf{Q}_{\mathrm{IEQ}}^{\mathrm{LS}}$, i.e., when computing AP, we
retain only the top 1000 positively weighted terms. However, when computing
cosine similarity between expanded queries and
\(\mathbf{Q}_{\mathrm{IEQ}}^{\mathrm{LS}}\), we use the full query vector
without truncation. Our intuition is that the negative weights provide
additional information about non-relevant terms and therefore contribute
meaningfully to the similarity estimation.
More details are discussed in Section~\ref{sec:discussion}.




\section{Results}
\label{results}
We now turn to the experimental results corresponding to the three
components of the framework presented in Section~\ref{proposed-framwork}:
\begin{enumerate*}[(1)]
\item constructing an ideal expanded query \(\mathbf{Q}_{\mathrm{IEQ}}\),
\item studying the correlation between Sim$(\mathbf{Q}_{\mathrm{exp}},
  \mathbf{Q}_{\mathrm{IEQ}})$ and \(\mathbf{Q}_{\mathrm{exp}}\)'s
  effectiveness, and
\item looking at the correlation between \(\mathbf{Q}_{\mathrm{exp}}\)'s
  effectiveness and how well it separates relevant and non-relevant documents.
\end{enumerate*}

\subsection{Retrieval effectiveness of \(\mathbf{Q}_{\mathrm{IEQ}}\) }
\label{results-ieq}
Table~\ref{tab:map_of_all_dataset} reports the MAP achieved by the ideal
queries constructed using the DFO and LS methods
(Sections~\ref{dfo-section} and \ref{ls-surrogate}). For context, we also
report the performance of $\mathbf{Q}_{\mathrm{Rocchio}}$, the inital
query vector constructed with full relevance judgements.
%
Unsurprisingly, the retrospectively tuned ideal queries yield very high
MAP, with the best scores on the Robust04 and DL 19-20 Passage collections being
near-perfect.
As discussed in Section~\ref{exp-settings}, the numbers of expansion terms
in \(\mathbf{Q}_{\mathrm{IEQ}}^{\mathrm{DFO}} \) and
\(\mathbf{Q}_{\mathrm{IEQ}}^{\mathrm{LS}}\) are different. For
completeness, therefore, we also report the MAP achieved when
\(\mathbf{Q}_{\mathrm{IEQ}}^{\mathrm{LS}}\) is truncated to the top 200
positively weighted terms. We observe that the two optimisation methods
are, broadly speaking, equally effective on all the collections, except
DL21-22 Passage, where DFO appears to yield better ideal queries than LS.

\begin{table*}[h!]
  \centering
  \caption{Retrieval effectiveness (MAP) achieved by the ideal queries
    generated using DFO and LS, on the Robust, DL19-20 Passage, DL19-20
    Document, and DL21-22 Passage collections. For reference, the initial
    Rocchio vector generated using full relevance assessments is also
    included. The highest MAP obtained for each collection is shown in
    bold. Superscript $\dagger$ denotes statistically significant difference with DFO, using t-test ($p$ < 0.05).}
  \begin{tabular}{l c c c c}
    \toprule
    \textbf{Method}     & \textbf{Robust} & \textbf{DL19-20 Passage} & \textbf{DL19-20 Document} & \textbf{DL21-22 Passage} \\
    \midrule
    $\mathbf{Q}_{\mathrm{Rocchio}}$                  &   0.6945 & 0.8096 & 0.6916 &  0.5383 \\\midrule
    
    \(\mathbf{Q}_{\mathrm{IEQ}}^{\mathrm{DFO}}\)                  &   0.8652 &  0.9240   & \textbf{0.8752}  & \textbf{0.8372} \\
    \(\mathbf{Q}_{\mathrm{IEQ}}^{\mathrm{LS}}(\lambda=0.1)\)   &  \textbf{0.9510$\dagger$} &  \textbf{0.9667$\dagger$}   &  0.8296$\dagger$ & 0.7456$\dagger$  \\
    \(\mathbf{Q}_{\mathrm{IEQ}}^{\mathrm{LS}}(\lambda=1)\)     &  0.8975$\dagger$ & 0.9520$\dagger$    & 0.8747 & 0.7415$\dagger$ \\\midrule
    \multicolumn{5}{c}{\textbf{Truncated to 200 terms}} \\\midrule
    \(\mathbf{Q}_{\mathrm{IEQ}}^{\mathrm{LS}}(\lambda=0.1)\) &  0.8757 & 0.9601   & 0.8241 & 0.7081    \\
    \(\mathbf{Q}_{\mathrm{IEQ}}^{\mathrm{LS}}(\lambda=1)\)  &  0.8186 & 0.9485   & 0.8558 &  0.7280 \\
    \bottomrule
  \end{tabular}
  \label{tab:map_of_all_dataset}
\end{table*}

\begin{figure*}[ht] 
  \centering
  \includegraphics[width=0.9\linewidth]{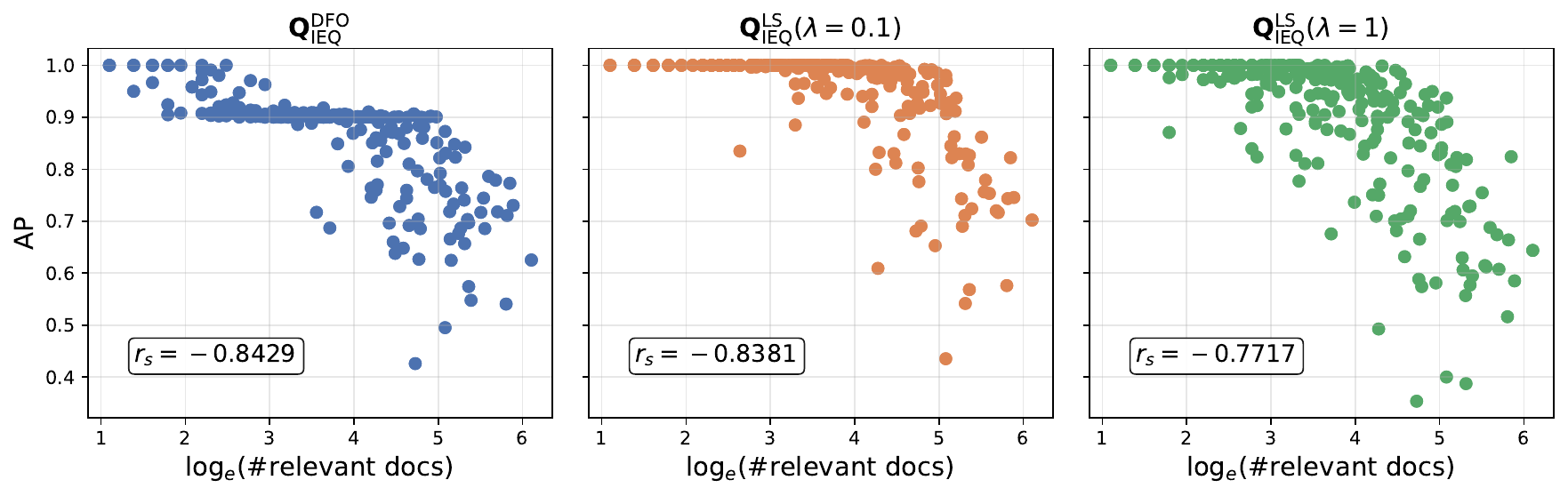}
  \caption{Per-query scatter plots of AP versus the number of relevant
    documents for three different ideal-query formulations on the Robust collection. The
    Spearman rank correlation coefficient $r_s$ between these two variables
    is reported in each plot.}
  \Description[AP vs number of reldocs plot]{Per-query scatter plots of AP
    versus the number of relevant documents for different ideal-query
    formulations on the Robust collection. The Spearman rank correlation
    coefficient $r_s$ between the two variables is reported in each plot.}
  \label{fig:main-ap-reldocs-robust}
\end{figure*} 


Figure~\ref{fig:main-ap-reldocs-robust} illustrates the relationship
between AP and the number of relevant documents for the three different
sets of ideal queries on the Robust collection. From the figure, it is
clear that many ideal queries generated by the LS method achieve a perfect
AP of 1.0. For DFO, there is an analogous dense band around AP=0.9, since
the DFO algorithm stops iterating when AP reaches 0.9 or above
(Section~\ref{exp-settings-ieq-dfo}). 

A second pattern that emerges is that ideal queries tend to achieve lower
AP values when the number of relevant documents is large. The Spearman
correlation between AP and the size of the relevant set across queries
shown in the figure indicate a strong negative correlation between these
two attributes. One possible explanation is that a large set of relevant
documents often encompasses multiple aspects or sub-topics, making it more
challenging for a single query representation to rank all relevant
documents highly. Similar trends are observed across the other collections;
corresponding figures are provided in
Appendix~\ref{additional-plots-ap-vs-relevant-docs}.
We observe that DL21-22 Passage collection contains a large number of judged relevant and non-relevant documents, and that the LS-based ideal queries achieve relatively low AP for these queries. Further figures are provided in Appendix~\ref{additional-plots-ap-vs-relevant-docs}.

Interestingly, while the DFO- and LS-based constructions both achieve high
MAP, they produce substantially different versions of
\(\mathbf{Q}_{\mathrm{IEQ}}\). Table~\ref{tab:avg_overlap_bw_ideal_queries}
reports the average per-query overlap in expansion terms between the ideal queries
(only the top 200 expansion
terms of \(\mathbf{Q}_{\mathrm{IEQ}}^{\mathrm{LS}}\) are considered in this
analysis).  Among the two LS variants, both the term overlap and cosine similarity are high. In contrast, the term overlap between DFO and LS is moderate, except for the DL21-22 Passage collection, where the overlap is higher, ranging from 65\% to 75\%.
This observation suggests that the ideal expanded query is not unique; there may
exist multiple distinct EQs that yield near-perfect retrieval performance.
Consequently, \(\mathbf{Q}_{\mathrm{IEQ}}\) should be viewed not as a
single optimal query, but rather as a family of highly effective query
representations. In future work, we would like to consider somewhat
modified versions of our research questions that consider a query's
proximity to \emph{any} member of this family of `ideal' representations,
rather than to a single, fixed ideal query.

\begin{table}[h!]
  \centering
  \caption{Average term overlap between ideal queries, 
    computed across all
    queries. For each ideal query variant, only the top 200 positively
    weighted terms are considered. Cosine similarity between the ideal queries is reported in square brackets.}
  %

  \resizebox{\linewidth}{!}{
    \begin{tabular}{lcccc}
      \toprule
      \textbf{Dataset} 
      & \textbf{Robust}            
      & \textbf{DL19-20 Passage}
      & \textbf{DL19-20 Document}  
      & \textbf{DL21-22 Passage} \\\midrule

      \(\mathbf{Q}_{\mathrm{IEQ}}^{\mathrm{DFO}}\) $\cap$ \(\mathbf{Q}_{\mathrm{IEQ}}^{\mathrm{LS}}(\lambda=0.1)\) 
      & 99.34 [0.5659]  & 115.71 [0.6477] & 89.33  [0.5315] & 130.33 [0.6416] \\

      \(\mathbf{Q}_{\mathrm{IEQ}}^{\mathrm{DFO}}\) $\cap$ \(\mathbf{Q}_{\mathrm{IEQ}}^{\mathrm{LS}}(\lambda=1)\)
      & 131.55 [0.6767]   & 133.52 [0.7755]   & 100.99 [0.6027]   & 153.67 [0.7421]   \\

      \(\mathbf{Q}_{\mathrm{IEQ}}^{\mathrm{LS}}(\lambda=0.1)\) $\cap$  \(\mathbf{Q}_{\mathrm{IEQ}}^{\mathrm{LS}}(\lambda=1)\)
      & 155.19 [0.8659] & 162.65 [0.9103] & 176.42 [0.9336] & 164.12 [0.8847]   \\
      
      \bottomrule
    \end{tabular}
  }
  \label{tab:avg_overlap_bw_ideal_queries}
\end{table}


\subsection{Correlation between Sim$(\mathbf{Q}_{\mathrm{exp}},
  \mathbf{Q}_{\mathrm{IEQ}})$ and $\mathit{AP}(\mathbf{Q}_{\mathrm{exp}})$}
\label{sec:synthetic}
To investigate our first research question \textbf{RQ1} --- \emph{for an
  EQ, how strongly does its similarity to the IEQ correlate with its
  retrieval effectiveness?} --- we generate a diverse set of synthetic EQs
in a controlled manner as described below, and measure the correlation
between their AP values and their similarity with (equivalently, distance
from) the ideal query.

The discussion in the next section (\ref{correlation-with-ideal-queries})
considers a large number of EQs produced by different existing QE approaches.
However, the AP values for these `real' EQs tend to vary within a
relatively small range (Table~\ref{tab:delta_ang}). For a more comprehensive analysis, we
need a more diverse set of EQs that spans a wider range of retrieval
effectiveness, and includes poor, moderate, and highly effective
expansions. In practice, obtaining this level of variation from existing QE
methods may be difficult. To overcome this limitation, we generate
synthetic expanded queries that exhibit a broader spectrum of retrieval
effectiveness.
We take the ideal query vector, randomly choose a plane passing through
this vector, and incrementally rotate the vector along this plane. We then
examine how AP varies as the vector is moved progressively away from the
original ideal query vector.

For this set of experiments, we use
\(\mathbf{Q}_{\mathrm{IEQ}}^{\mathrm{LS}}(\lambda=0.1)\) as the ideal
query. 
We rotate the IEQ
vector 
by $\{\theta, 2\theta, 3\theta, \ldots, 100\theta\}$, where $100\theta =
90^{\circ}$.\footnote{We observe that, when the IEQ vector is rotated by
  $90^{\circ}$, the resulting vector yields AP=0 for all the queries across
  all the collections, confirming that the generated set spans a wide range
  of AP values.} The vector obtained at each of these 100 steps
corresponds to a synthetic EQ. For retrieval, we retain only the top 200
positively weighted terms of each synthetic query and compute its AP. We
also note the cosine similarity between each synthetic EQ vector and the
original IEQ vector. Finally, we measure the correlation for the set of
(\text{similarity}, AP) pairs thus obtained.

\begin{table}[h!]
  \centering
  \caption{Mean Pearson and Spearman correlation coefficients between the
    similarity of synthetically generated expanded queries to the ideal
    query \(\mathbf{Q}_{\mathrm{IEQ}}^{\mathrm{LS}}(\lambda=0.1)\), and
    their retrieval effectiveness, across all collections.}
  \resizebox{0.5\linewidth}{!}{
    \begin{tabular}{lcc}
      \toprule
      \textbf{Dataset} & Pearson $\rho$ & Spearman $r_s$ \\
      \midrule
      Robust              & 0.6783 & 0.4497  \\
      DL19-20 Passage     & 0.8823 & 0.8349  \\
      DL19-20 Document    & 0.7749 & 0.5311 \\
      DL21-22 Passage     & 0.8834 & 0.8512 \\
      \bottomrule
    \end{tabular}
  }
  \label{tab:synthetic_query_corr}
\end{table}

\begin{figure*}[h!]
    \centering
    \includegraphics[width=0.6\linewidth]{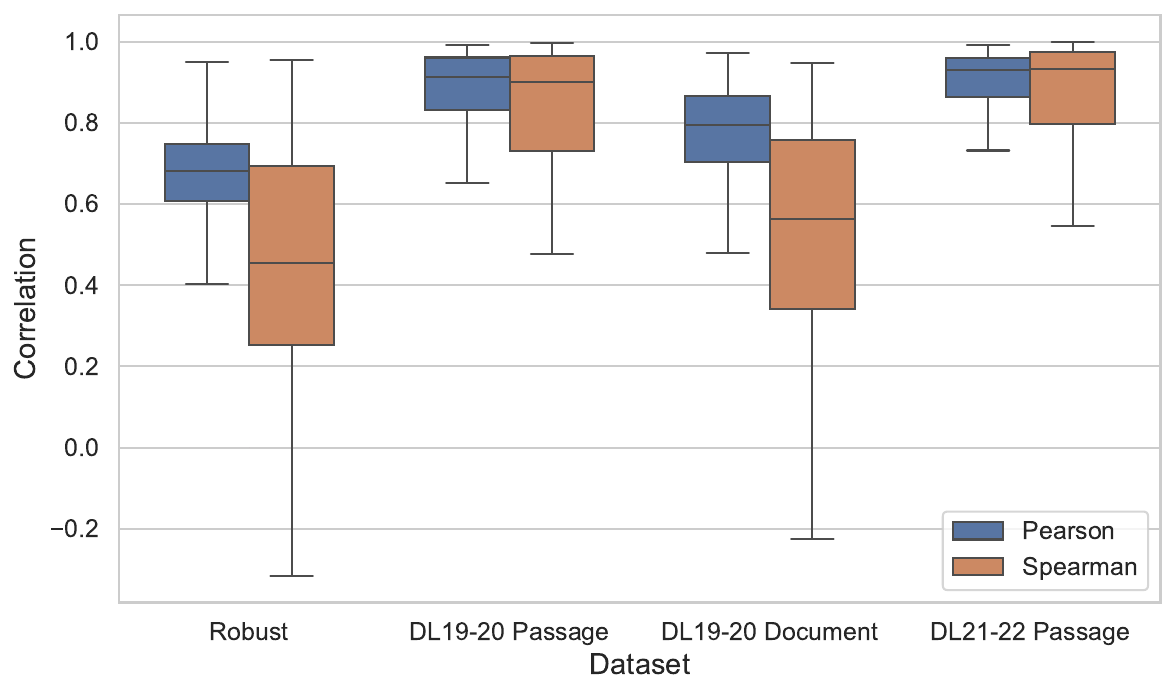}
    \caption{Box plot of Pearson and Spearman correlations between the similarity of synthetically generated expanded queries to the ideal query \(\mathbf{Q}_{\mathrm{IEQ}}^{\mathrm{LS}}(\lambda=0.1)\) and their retrieval effectiveness across all collections.}
    \Description[box plot for synthetic queries]{Box plot of Pearson and Spearman correlations across the four different collections.}
    \label{fig:boxplot-synthetics-queries}
\end{figure*}




Table~\ref{tab:synthetic_query_corr} reports the mean correlation between
Sim$(\mathbf{Q}_{\mathrm{exp}}, \mathbf{Q}_{\mathrm{IEQ}})$ and
$\mathit{AP}(\mathbf{Q}_{\mathrm{exp}})$. The consistently high values
across different collections indicate a strong relationship between these
two attributes. 
These findings provide strong evidence in support of an affirmative answer
to \textbf{RQ1}, suggesting that proximity to the ideal query is highly
predictive of retrieval performance.

\begin{figure}
	\centering
		\includegraphics[width=0.7\linewidth]{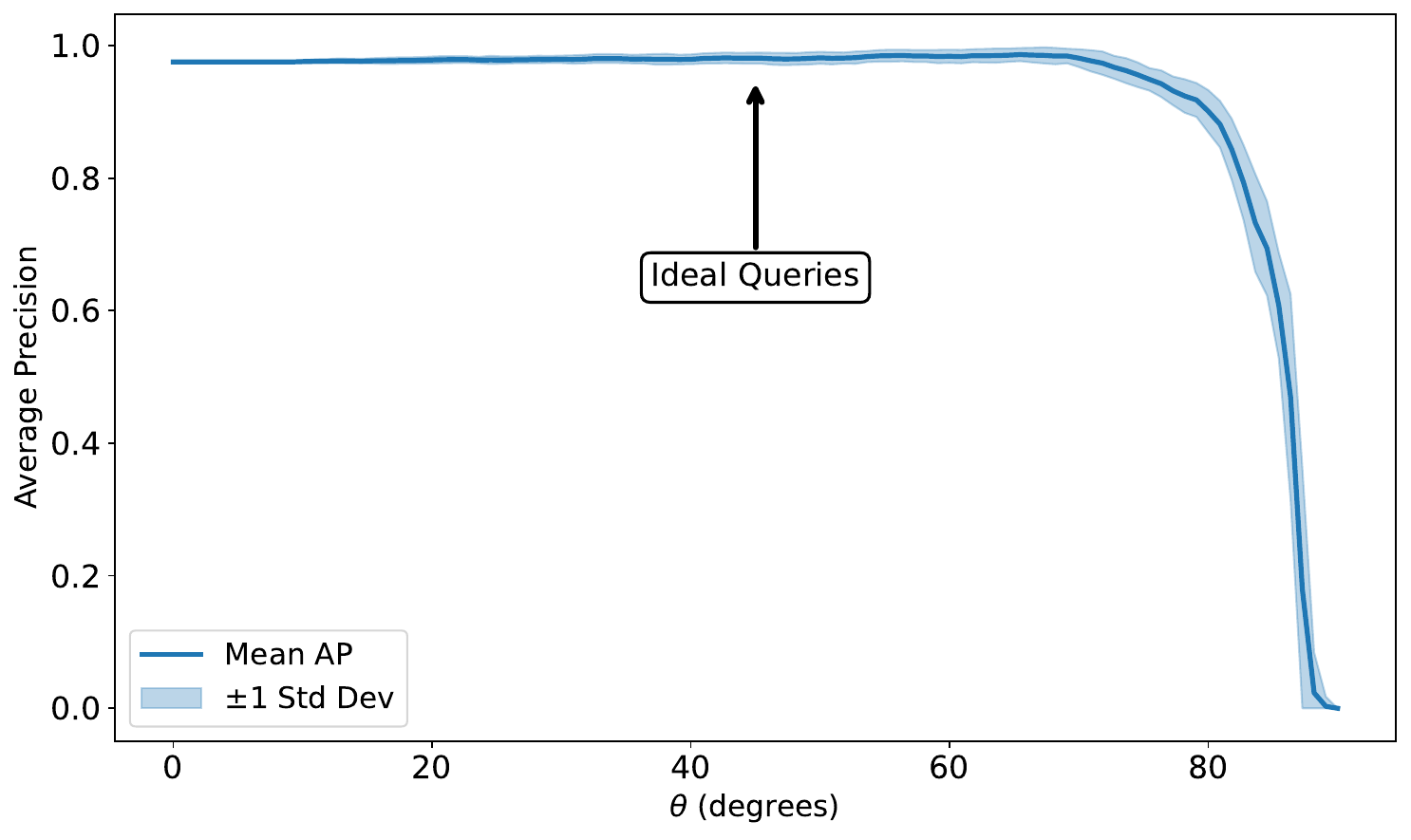}
	\caption{Y-axis shows the AP variation and the X-axis plots the angle from the ideal query vector for the synthetically generated versions of the query \emph{`black bear attacks' (Robust qid: 336)}.}
	\label{fig:synthetic-query}
    \Description[ap vs theta]{Y-axis shows the AP variation and the X-axis plots the angle from the ideal query vector for the synthetically generated expansion terms of the query \emph{`black bear attacks' (Robust qid: 336)}.}
\end{figure}

Figure~\ref{fig:boxplot-synthetics-queries} presents more details about the
distribution of the Pearson and Spearman correlation coefficients across
queries. On the Robust and DL19-20 Document collections, a few queries
exhibit negative Spearman correlations. A closer examination reveals that,
for most of these queries, AP remains nearly unchanged as the query is
rotated up to approximately 60-70$^{\circ}$, after which it sometimes
\emph{increases} slightly. Figure~\ref{fig:synthetic-query} shows such an
example from the Robust dataset (query 336: ``black bear attacks''). The
figure suggests that many ideal queries lie within the 70$^{\circ}$ region.
Additional investigation of these anomalous queries is left for future
work.

\subsection{Correlation for `real' expanded queries}
\label{correlation-with-ideal-queries}
Having established the relationship between Sim$(\mathbf{Q}_{\mathrm{exp}},
\mathbf{Q}_{\mathrm{IEQ}})$ and $\mathit{AP}(\mathbf{Q}_{\mathrm{exp}})$ in
a controlled synthetic setting, we next measure the correlation between
these two quantities for real EQs generated by existing QE methods. We
follow the method described in Section~\ref{pairwise-listwise-corr-define}
to compute mean Pearson ($\rho$) and Spearman ($r_s$) correlation
coefficients.
Table~\ref{tab:corr_results} reports $\rho$ and $r_s$ across all datasets,
for three different ideal query formulations. Reasonable correlation (0.40
or higher\footnote{Correlations of a similar magnitude have been reported in related  IR studies, like Query Performance Prediction~\cite{bertqpp, qpp_tois}, although the underlying prediction tasks and experimental settings differ from those considered here.}) is observed in all cases, suggesting once again that there is a
positive relation between proximity to the ideal query and retrieval
performance (although this correlation is not as strong as it was for
synthetic queries). From a closer look at Table~\ref{tab:corr_results}, the
following patterns emerge.
\begin{itemize}
\item When LS-based ideal queries are used, the correlation coefficients
  for the Robust and DL21-22 Passage collections are noticeably higher.
  For the DL19-20 collections (both document and passage), the values are
  more moderate.
\item For the passage collections, changing $\lambda$ from 0.1 to 1.0 has a
  noticeable impact on $\rho$ and $r_s$; however, the choice of $\lambda$
  does not seem to matter much for the document collections.
\item When using the DFO-based ideal queries, good correlation is observed
  for the DL21-22 Passage collection, but the correlation is moderate
  (0.4--0.5) for all the other collections. DL21-22 Passage is also the
  only collection where DFO-based ideal queries yield better correlation
  than LS-based ideal queries; on all other collections LS-based ideal
  queries consistently yield stronger correlations.

\end{itemize}
%



\begin{table*}[h!]
\centering
\caption{Mean Pearson and Spearman correlation coefficients between the similarity of real expanded queries to different ideal query formulations and the AP of the expanded queries on the Robust, DL19-20 Passage, DL19-20 Document, and DL21-22 Passage collections.}
\begin{tabular}{l cc cc cc cc}
\toprule
\textbf{Method}     & \multicolumn{2}{c}{\textbf{Robust}}  & \multicolumn{2}{c}{\textbf{DL19-20 Passage}} &  \multicolumn{2}{c}{\textbf{DL19-20 Document}}  & \multicolumn{2}{c}{\textbf{DL21-22 Passage}}\\
\cmidrule(lr){2-3} \cmidrule(lr){4-5} \cmidrule(lr){6-7} \cmidrule(lr){8-9}

     & $\rho$ & $r_s$ & $\rho$ & $r_s$ & $\rho$ & $r_s$ & $\rho$ & $r_s$ \\

\midrule
\(\mathbf{Q}_{\mathrm{IEQ}}^{\mathrm{DFO}}\)                   & 0.4366  & 0.4185 & 0.4762 & 0.4488 & 0.4245 & 0.4030 & 0.6774 & 0.5413   \\
\(\mathbf{Q}_{\mathrm{IEQ}}^{\mathrm{LS}}(\lambda=0.1)\)   & 0.5997 & 0.5770  & 0.4110 & 0.4097 & 0.4260 & 0.4390  & 0.4916 & 0.4360 \\
\(\mathbf{Q}_{\mathrm{IEQ}}^{\mathrm{LS}}(\lambda=1)\)     &  0.6053 & 0.5716  & 0.5078 & 0.4965 & 0.4680 & 0.4685  &  0.6395  & 0.5459  \\\midrule


\bottomrule
\end{tabular}
\label{tab:corr_results}
\end{table*}

\begin{table}[h!]
\centering
\caption{Mean correlation (Pearson's $\rho$ and Spearman's $r_s$) between the Separability measure and actual AP across all expanded queries on the Robust, DL19-20 Passage, DL19-20 Document, and DL21-22 Passage collections.}
\resizebox{0.5\linewidth}{!}{
\begin{tabular}{l c c}
\toprule
\textbf{Dataset} 
& Pearson's $\rho$ & Spearman's $r_s$  \\ 
\midrule
Robust              & 0.7849 & 0.7359 \\ 
DL19-20 Passage    & 0.6504 & 0.6417 \\ 
DL19-20 Document   & 0.6503 & 0.6241 \\ 
DL21-22 Passage    & 0.7024 & 0.5971 \\ 
\bottomrule
\end{tabular}
}
\label{tab:corr_ap_vs_separability}
\end{table}


\subsection{Correlation between separability and performance}
\label{separability-correlation-results}
We next examine the results pertaining to \textbf{RQ2} --- how strongly
does an EQ's performance correlate with its ability to separate relevant
and non-relevant documents? As discussed in
Section~\ref{separability-formulation}, we consider separability in terms
of BM25 scores (Equation~\ref{qe-score-separability-final}). Only judged
relevant and non-relevant documents are considered in these experiments, and
separability is measured using \textit{Cohen's} \(d\).

Table~\ref{tab:corr_ap_vs_separability} reports the mean correlation
between separability and AP of the expanded queries.  
We observe that the correlation
between AP
and the proposed separability measures is generally much stronger than its
correlation with the retrieval effectiveness of real EQs (reported in the
preceding section). The strongest correlation ($\rho$ = 0.7849) is obtained
on the Robust collection. Thus, \textbf{RQ2} also seems to have a positive
answer, i.e., if the scores of relevant and non-relevant documents are
well-separated for an EQ, it is likely to achieve good retrieval effectiveness.
This relationship is intuitive, as achieving high AP generally requires relevant documents to be ranked ahead of non-relevant documents, which can also result in greater separation between their scores. However, the correlation is not perfect, indicating that high AP does not necessarily imply high separability. In particular, some EQs may achieve good AP despite exhibiting relatively low separation between the scores of relevant and non-relevant documents. A more detailed analysis of these cases, including the concordance pairs, is left for future work.

\subsection{Explaining poor performance of different queries}
\label{sec:explain-poor-performance}
Thus far, we have focused mainly on explaining the difference in
performance across EQs generated for a single query by different QE
algorithms and different parameter settings for these algorithms. We now
extend our experiments to consider performance variation across different
queries.\footnote{Appendix~\ref{variability} provides supplementary context
  for the results reported here.} We are particularly interested in queries
for which automatic QE techniques are generally ineffective. As before, we
investigate whether this variation (or generally poor performance) can be
explained by the overall proximity of their respective EQs to the
respective ideal queries.

Intuitively, a query may be considered hard for QE techniques if a diverse
set of QE methods consistently fails to improve its retrieval
effectiveness. We may consider two aggregation strategies to represent the
general effectiveness of QE techniques for this query: average performance
and maximum performance, as defined in the equation below. The maximum
performance setting captures the intuition that a query is hard when even
its best performing expanded variant achieves low retrieval effectiveness,
indicating that automatic QE approaches struggle to improve the query.
\begin{equation*}
  AP_{\mathrm{avg}}(Q)
  =
  \frac{1}{k}\sum_{j=1}^{k} AP\left(\mathbf{Q}_{\mathrm{exp}}^{(j)}\right),
  \qquad\qquad
  AP_{\mathrm{max}}(Q)
  =
  \max_k AP\left(\mathbf{Q}_{\mathrm{exp}}^{(k)}\right).
  \label{eq:ap-aggregation}
\end{equation*}
Here, $\mathit{AP}(\mathbf{Q}_{\mathrm{exp}}^{(1)}),
\mathit{AP}(\mathbf{Q}_{\mathrm{exp}}^{(2)})$, $\dots,$
$\mathit{AP}(\mathbf{Q}_{\mathrm{exp}}^{(k)})$, denote the AP values achieved
by the $k$ expanded variants of an initial query $Q$. We use the same
aggregation methods to compute an overall proximity of the set of EQs to
the ideal query \(\mathbf{Q}_{\mathrm{IEQ}}\).
\begin{equation*}
  \operatorname{Sim}_{\mathrm{avg}}(Q)
  =
  \frac{1}{k}\sum_{j=1}^{k}
  \operatorname{Sim}\left(
    \mathbf{Q}_{\mathrm{exp}}^{(j)},\mathbf{Q}_{\mathrm{IEQ}}
  \right),
\end{equation*}

\begin{equation*}
  \operatorname{Sim}_{\mathrm{max}}(Q)
  =
  \operatorname{Sim}\left(
    \mathbf{Q}_{\mathrm{exp}}^{(k^*)},\mathbf{Q}_{\mathrm{IEQ}}
  \right),
  \qquad\qquad
  k^*=\arg\max_k AP\left(\mathbf{Q}_{\mathrm{exp}}^{(k)}\right).
  \label{eq:similarity-aggregation}
\end{equation*}
Finally, for each aggregation strategy $A \in
\{\mathrm{avg},\mathrm{max}\}$, we compute the Pearson correlation between
AP and similarity:
\begin{equation*}
\operatorname{corr}\left(
\{AP_A(Q_i)\}_{i=1}^{N},
\{\operatorname{Sim}_A(Q_i)\}_{i=1}^{N}
\right),
\end{equation*}
where $N$ denotes the total number of queries in a collection.

\begin{table*}[h!]
  \centering
  \caption{Pearson correlation between aggregate AP and similarity for
    expanded queries. Three different ideal query representations are
    considered. For each collection, correlations are reported for the
    average and maximum similarity. The highest correlation within each
    collection and aggregation setting is shown in bold.}
  \label{tab:qpp}
  \begin{tabular}{lcc cc cc cc}
    \toprule
    & \multicolumn{2}{c}{\textbf{Robust}}
    & \multicolumn{2}{c}{\textbf{DL19-20 Passage}}
    & \multicolumn{2}{c}{\textbf{DL19-20 Document}}
    & \multicolumn{2}{c}{\textbf{DL21-22 Passage}} \\
    \cmidrule(lr){2-3}
    \cmidrule(lr){4-5}
    \cmidrule(lr){6-7}
    \cmidrule(lr){8-9}
    \textbf{Method}
    & Avg. & Max.
    & Avg. & Max.
    & Avg. & Max.
    & Avg. & Max. \\
    \midrule

    $\mathbf{Q}_{\mathrm{IEQ}}^{\mathrm{DFO}}$ & 0.3790 & 0.4608 & 0.6770 & 0.6693 & 0.4146 & 0.3556 & \textbf{0.5043} & 0.5378 \\
    $\mathbf{Q}_{\mathrm{IEQ}}^{\mathrm{LS}}(\lambda=0.1)$ & 0.7393 & 0.7150 & 0.7379 & 0.6344 & 0.4547 & 0.4507 & 0.3059 & 0.2698 \\
    $\mathbf{Q}_{\mathrm{IEQ}}^{\mathrm{LS}}(\lambda=1)$ & \textbf{0.7421} & 0.6915 & \textbf{0.7656} & 0.6400 & \textbf{0.4961} & 0.4647 & 0.3811 & 0.3617 \\

    \bottomrule
  \end{tabular}
\end{table*}

Table~\ref{tab:qpp} reports the correlation between aggregated AP and
similarity to three ideal query representations across all collections. We
observe moderate to strong correlations across all collections when using
\(\mathbf{Q}_{\mathrm{IEQ}}^{\mathrm{LS}} (\lambda=1)\), except for the
DL21-22 Passage collection, where the correlations are relatively weak. The
correlations obtained using
\(\mathbf{Q}_{\mathrm{IEQ}}^{\mathrm{LS}}(\lambda=0.1)\) are slightly
lower. 
%
Overall, these findings provide evidence supporting a positive answer to
\textbf{RQ3}: similarity to the ideal query helps explain why automatic QE
techniques are less effective for some queries. Queries whose expanded
variants are close to the ideal query tend to achieve higher retrieval
effectiveness, whereas those whose expanded variants lie farther from it
tend to achieve lower effectiveness.


\section{Discussion: explaining variations in the similarity--effectiveness relationship}
\label{sec:discussion}
Although our framework suggests a positive relationship between similarity
to the ideal query and retrieval effectiveness on average, the observed
correlations are moderate for some collections and queries. In this
section, we investigate the reasons for these weaker correlations. 

To facilitate this analysis, we introduce a restricted evaluation setting.
For any query, we limit the evaluation corpus to only the documents that
have been explicitly judged for that query. Thus, given a query, we
exhaustively compute BM25 scores for all judged documents, and rank only
these documents. We define \emph{Restricted AP} as the Average Precision
computed over this restricted ranking.

Computing standard, collection-wide AP, on the other hand, requires ranking
documents from the entire collection. Top-$k$ retrieval algorithms
generally used by IR systems like Lucene make use of optimisations that
assume that a document's score cannot decrease as more query terms are
considered (\url{https://issues.apache.org/jira/browse/LUCENE-7996}). This
assumption does not hold when query terms can have negative weights. Thus,
efficiently computing collection-wide AP requires truncating synthetic
queries to their top-\emph{k} positively weighted terms. This step discards
negatively weighted terms (``bad'' terms that occur preferentially in
non-relevant documents, and are therefore detrimental to effectiveness)
from query vectors, and prevents us from accounting for such bad terms when
calculating Sim$(\mathbf{Q}_{\mathrm{exp}}, \mathbf{Q}_{\mathrm{IEQ}})$.

When computing Restricted AP, however, we work with a comparatively small
set of judged documents. This makes it feasible to compute a document
ranking in the presence of negative term weights, as is the case for
\(\mathbf{Q}_{\mathrm{IEQ}}^{\mathrm{LS}}\) and synthetic queries derived
from it.

Below, we first show that Restricted AP is a good approximation of
collection-wide AP. We then use this setting to reexamine how retrieval
effectiveness changes as queries move away from the ideal query, and to
explain why correlations observed for real EQs are sometimes weaker than
those obtained in the synthetic-query setting.

\subsection{AP vs. Restricted AP of \(\mathbf{Q}_{\mathrm{exp}}\)}
\label{ap-vs-restricted}
To validate whether Restricted AP serves as a reliable proxy for standard
AP, we compute both metrics for each of the 90 QE
configurations.\footnote{Standard AP can be readily calculated for these `real'
  queries since they are generated by existing QE methods that do not
  assign negative weights.} For every query, we then measure the
correlation between the resulting AP and Restricted AP across all
configurations, and average these correlations over all queries.
Table~\ref{tab:corr_ap_vs_restricted_ap} reports the results.
We observe consistently high Pearson and Spearman correlations across all
collections, indicating that Restricted AP preserves the relative
effectiveness of QE methods while substantially reducing the computational
cost of evaluation.
We therefore use Restricted AP in place of AP in the following analyses to
better understand the similarity-effectiveness relationship.

\begin{table}[h!]
\centering
\caption{Correlation (Pearson's $\rho$ and Spearman $r_s$) between
  restricted AP and actual AP across all expanded variants of a query,
  averaged over all queries in a collection.}
\resizebox{0.5\linewidth}{!}{
\begin{tabular}{l c c}
\toprule
\textbf{Dataset}    & $\rho$ & $r_s$ \\ 
\midrule
Robust              & 0.9689 & 0.9448 \\ 
DL19-20 Passage     & 0.8560 & 0.8263 \\ 
DL19-20 Document    & 0.8252 & 0.8203 \\ 
DL21-22 Passage    & 0.7774 & 0.6767  \\ 
\bottomrule
\end{tabular}
}
\label{tab:corr_ap_vs_restricted_ap}
\end{table}

\subsection{Restricted AP vs. Similarity to \(\mathbf{Q}_{\mathrm{IEQ}}\)}
\begin{figure}
  \centering
  \includegraphics[width=0.6\linewidth]{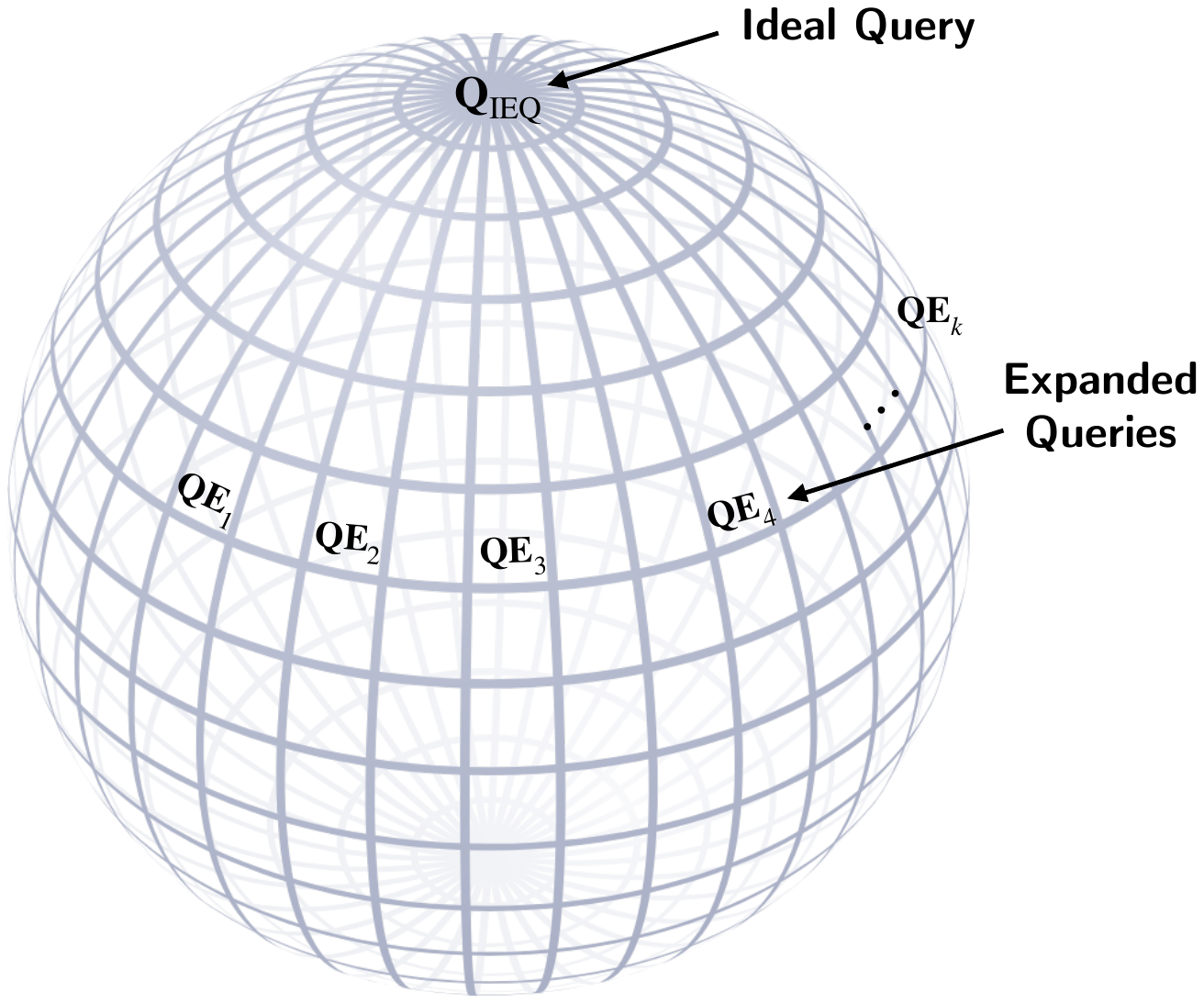}
  \caption{Trajectories along the great circles constructed from
    \(\mathbf{Q}_{\mathrm{IEQ}}\) and each \(\mathbf{Q}_{\mathrm{exp}}\),
    along which synthetic queries are generated.} 
  \Description[great circle trajectories]{Trajectories along the great
    circles constructed from \(\mathbf{Q}_{\mathrm{IEQ}}\) and each
    \(\mathbf{Q}_{\mathrm{exp}}\), along which synthetic queries are
    generated.}
  \label{fig:unit-sphere}
\end{figure}

Using Restricted AP, we now revisit the question of how retrieval
effectiveness changes as EQs move away from the ideal query. Recall that,
in Section~\ref{sec:synthetic}, we constructed synthetic queries by
progressively rotating the IEQ vector in a \emph{randomly} chosen plane
passing through it. In this section, for a given query ${Q}$, we consider
each expanded version \(\mathbf{Q}_{\mathrm{exp}}\) in turn, and
progressively rotate \(\mathbf{Q}_{\mathrm{IEQ}}\) through a full
\(360^\circ\) in the plane defined by \(\mathbf{Q}_{\mathrm{exp}}\) and
\(\mathbf{Q}_{\mathrm{IEQ}}\), instead of a randomly chosen plane. At 500
regularly spaced points along this trajectory, we take the rotated version
of \(\mathbf{Q}_{\mathrm{IEQ}}\) as our synthetic query vector, and compute
its similarity to \(\mathbf{Q}_{\mathrm{IEQ}}\) together with the
corresponding Restricted AP.\footnote{Note that evaluating collection-wide
  AP for the set of 500 synthetic queries generated along each great-circle
  trajectory would be computationally very expensive, as it would require
  retrieving and ranking documents from the entire collection.}
Figure~\ref{fig:unit-sphere} illustrates this process using an ideal query
and various different expanded queries.

\subsubsection*{Observations.}
\begin{figure*}[t]
\centering

\begin{subfigure}{0.48\textwidth}
    \includegraphics[width=\linewidth]{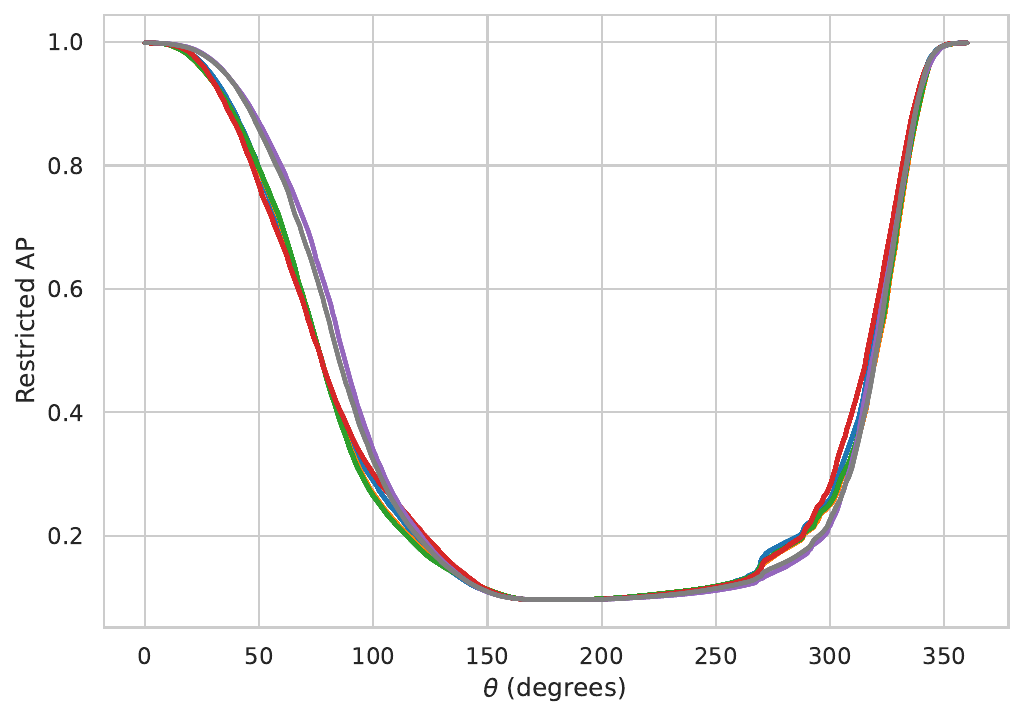}
    \caption{DL19-20 Passage}
\end{subfigure}
\hfill
\begin{subfigure}{0.48\textwidth}
    \includegraphics[width=\linewidth]{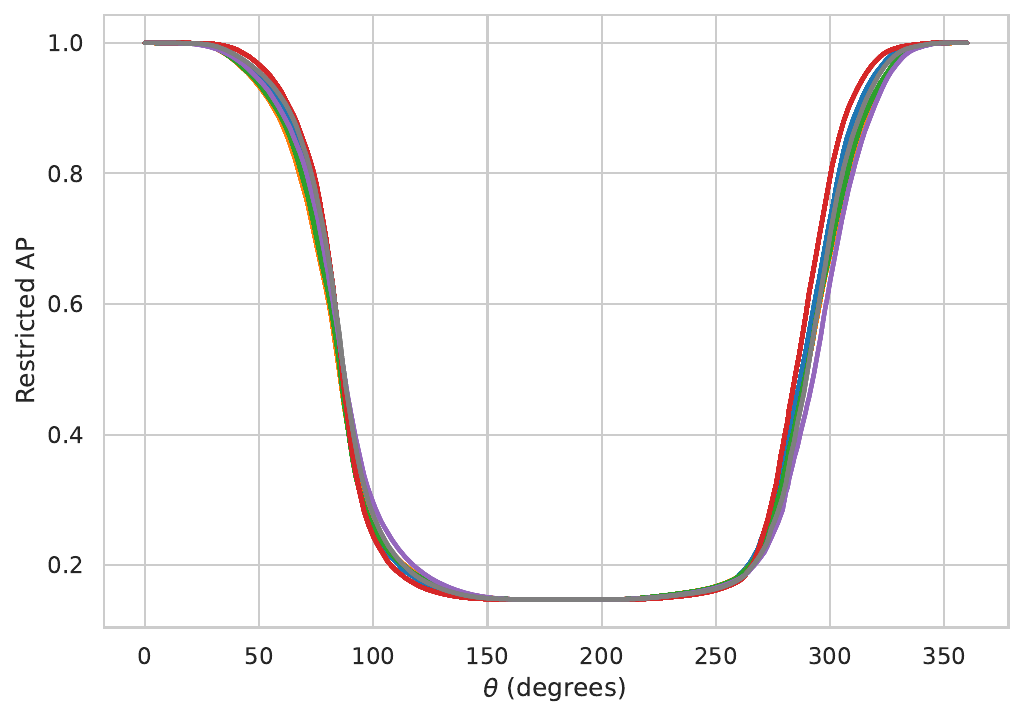}
    \caption{DL19-20 Document}
\end{subfigure}
\begin{tabular}{cccccc}
{\color{rm3}$\bullet$} RM3 & 
{\color{bo1}$\bullet$} Bo1 &
{\color{kl}$\bullet$} KL &
{\color{ceqe}$\bullet$} CEQE &
{\color{hyde}$\bullet$} HyDE & 
{\color{hyderocchio}$\bullet$} HyDE-Rocchio
\end{tabular}
\caption{Variation in Restricted AP along great circles originating from $\textbf{Q}_\text{IEQ}^\text{LS}$ $(\lambda=0.1)$ and passing through 500 synthetic queries generated from expanded queries produced by RM3, Bo1, KL, CEQE, HyDE, and HyDE-Rocchio.}
\label{fig:restricted_ap_vs_theta}
\end{figure*}

Figure~\ref{fig:restricted_ap_vs_theta} illustrates the relationship
between Restricted AP and the angular distance of the synthetic queries
from the ideal query (equivalent to Sim$(\mathbf{Q}_{\mathrm{exp}},
\mathbf{Q}_{\mathrm{IEQ}})$) for the DL19-20 Passage and Document
collections. For a clean visual representation, we draw a single curve for
each of the six QE algorithms used in our experiments
(Section~\ref{qe-approaches-explored}), by averaging across all queries in
the collection, as well as across all parameter variations for that
algorithm. It is clear that synthetic queries that are more similar to the
ideal query consistently achieve higher Restricted AP, while queries
farther from \(\mathbf{Q}_{\mathrm{IEQ}}\) exhibit lower effectiveness. A
similar phenomenon is observed for the other two collections. This
observation is consistent with the strong correlations reported in
Section~\ref{sec:synthetic}.

A closer look at Figure~\ref{fig:restricted_ap_vs_theta} also provides
insights into why the correlations observed for real EQs are moderate for
some collections
(Table~\ref{tab:corr_results}). 
In the synthetic setting, queries are generated along a single great-circle
trajectory originating from \(\mathbf{Q}_{\mathrm{IEQ}}\). Along such a
trajectory, the weights of terms in the synthetic queries deviate
systematically from their weights in \(\mathbf{Q}_{\mathrm{IEQ}}\). The
relationship between similarity and Restricted AP is thus highly
structured, resulting in strong correlations.
In contrast, real EQs correspond to points lying on different
trajectories (great circles) emanating from \(\mathbf{Q}_{\mathrm{IEQ}}\).
Although a positive relationship between similarity and effectiveness holds
along each individual trajectory, the shape and strength of this
relationship may differ across trajectories. Consequently, when EQs from
different trajectories are analyzed together (as in
Section~\ref{correlation-with-ideal-queries}), these trajectory-specific
variations weaken overall correlation, even when the within-trajectory
relation remains strong. This explains the relatively weaker correlations
observed for real EQs on the DL19-20 Document collection
(Table~\ref{tab:corr_results}).

\begin{figure}
  \centering
  \includegraphics[width=0.6\linewidth]{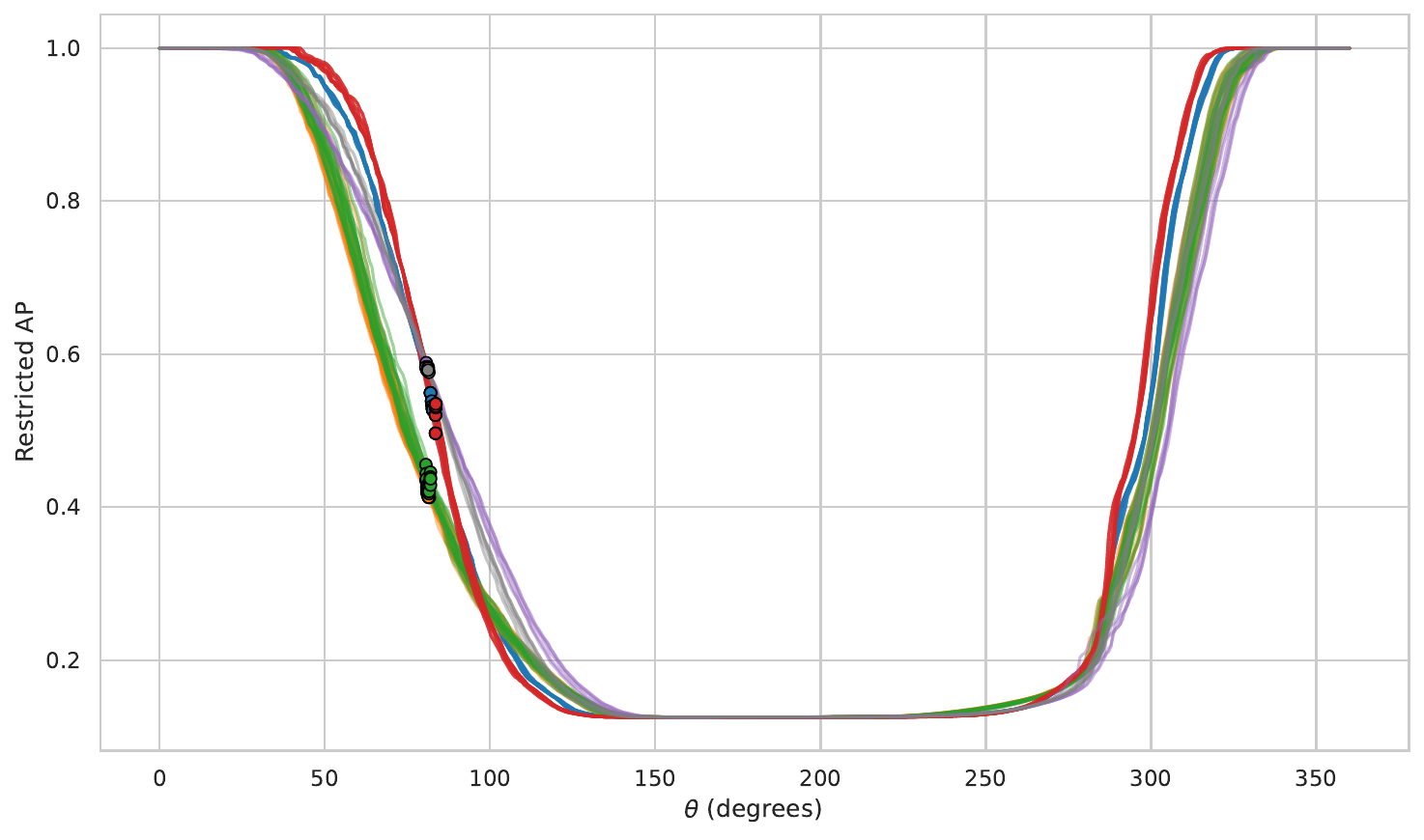}
  \caption{Query-specific variation (only for qid 730539 of the DL19-20
    Document collection) of Restricted AP along 90 great circles
    (originating from $\textbf{Q}_\text{IEQ}^\text{LS}$ $(\lambda=0.1)$ and
    passing through 90 real expanded versions, generated using RM3, Bo1,
    KL, CEQE, HyDE, and HyDE-Rocchio methods, at different parameter
    settings).}
  \label{fig:restricted_ap_vs_theta_single_q}
  \Description[ap vs theta]{The angular deviation from the ideal query
    vector for the synthetically generated EQs is plotted along the X-axis,
    while the Y-axis corresponds to Restricted AP}
\end{figure}

Figure~\ref{fig:restricted_ap_vs_theta_single_q} demonstrates this problem
for a single query (qid 730539 of the DL19-20 Document collection:
\emph{what is chronometer who invented it}). The figure contains 90 lines,
each of which shows the variation of Restricted AP as one ideal version of
this query ($\textbf{Q}_\text{IEQ}^\text{LS}$ for $\lambda=0.1$) is rotated
through a great circle formed by the ideal query itself, and one of the 90
real expanded variants of the original query. The circled points on the
lines correspond to
the real EQs themselves. As mentioned above, the within-trajectory relation
remains strong for each of the 90 lines; however, across trajectories,
there are many pairs $(EQ_i,EQ_j)$ of circled points (i.e., real EQs) such
that \emph{both} the angular deviation and Restricted AP are higher for
$EQ_i$ than for $EQ_j$. This naturally leads to low values of the
correlation coefficients (for this query, $\rho = -0.42049$ and $r_s =
-0.3151$).

\subsection{Angular spread of expanded versions of a query}
\begin{figure*}
    \centering
    \includegraphics[width=1\linewidth]{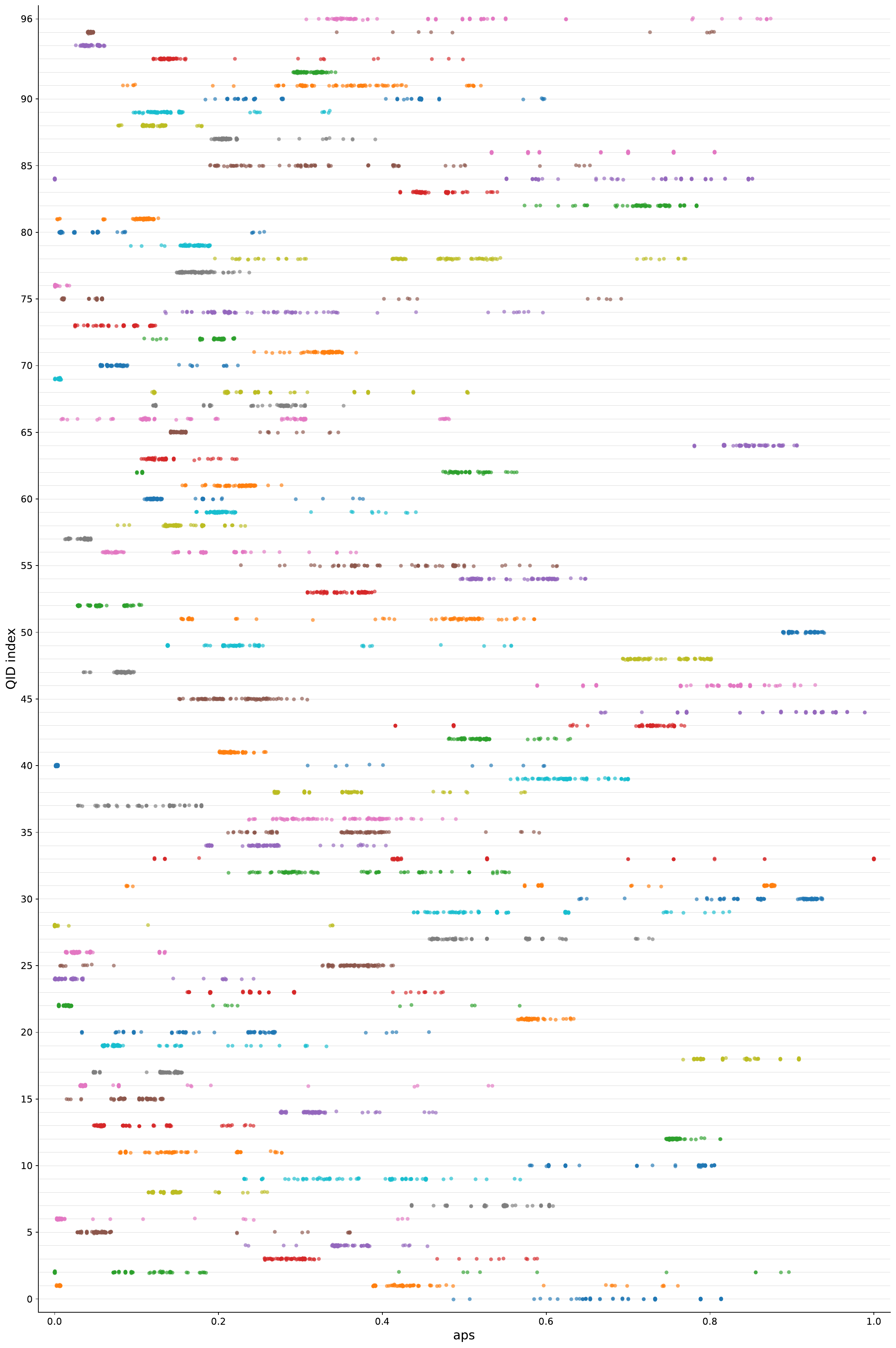}
    \caption{Per-query variation in AP across different expanded queries (RM3, Bo1, KL, CEQE, HyDE and HyDE-Rocchio) for the benchmark queries in the DL19--20 Passage collection.}
    \label{fig:ap-variation-all-qs}
\end{figure*}

\begin{figure*}
    \centering
    \includegraphics[width=0.9\linewidth]{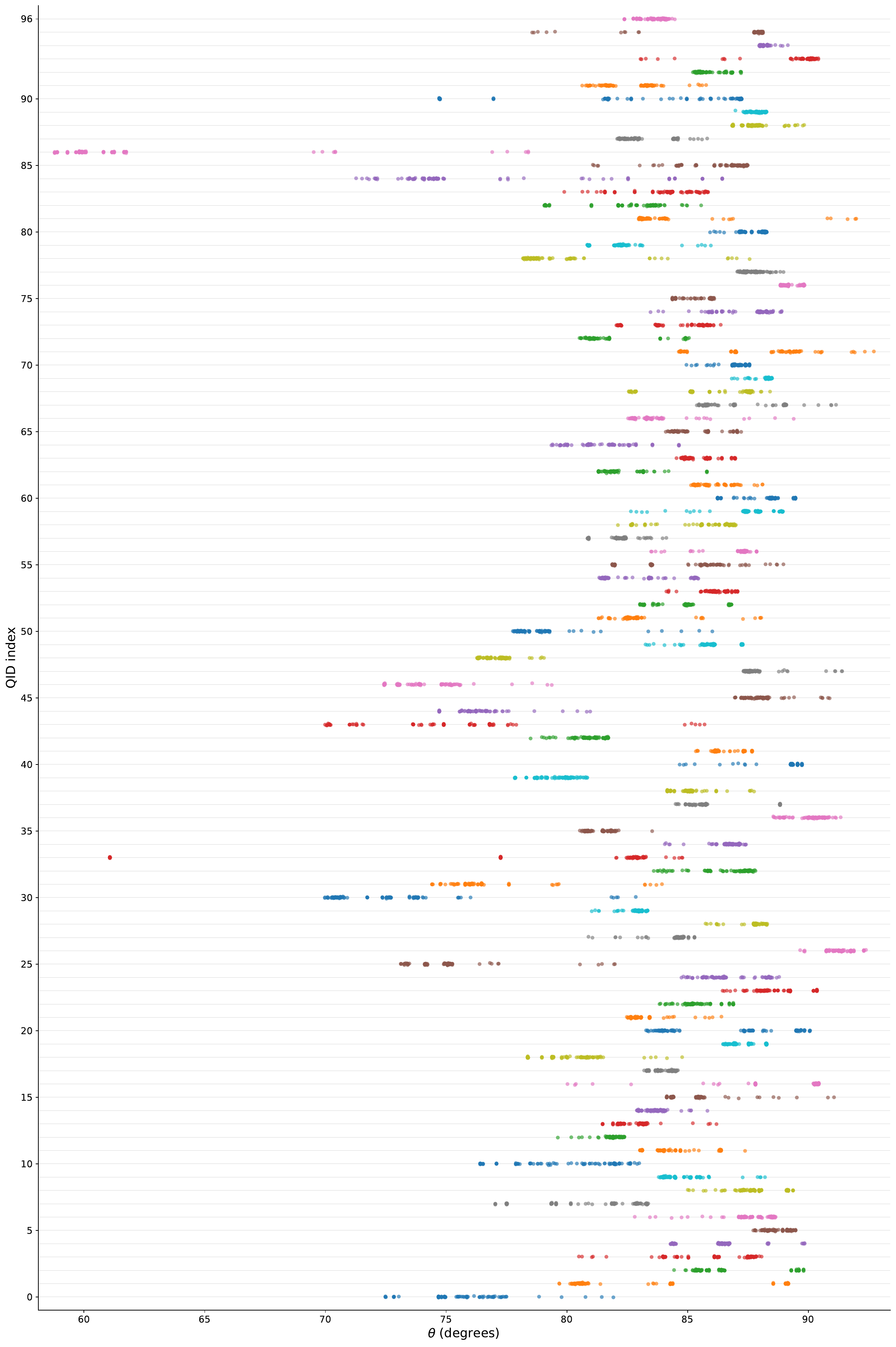}
    \caption{Per-query variation in angular distance from \(\mathbf{Q}_{\mathrm{IEQ}}^{\mathrm{LS}}(\lambda=1)\) across different expanded queries (RM3, Bo1, KL, CEQE, HyDE and HyDE-Rocchio) for the benchmark queries in the DL19--20 Passage collection.}
    \label{fig:angle-variation-all-qs}
\end{figure*}

Another pattern emerges from
Figure~\ref{fig:restricted_ap_vs_theta_single_q}: in terms of horizontal
spread, the circled points (corresponding to real EQs) span a very narrow
region. In other words, the angular distances of the EQs from the IEQ lie
within a small range (in Figure~\ref{fig:restricted_ap_vs_theta_single_q},
the actual range is $80.5^{\circ}$-$83.5^{\circ}$). This narrow angular
spread suggests an explanation for the weak or even negative correlations
observed for some queries. Table~\ref{tab:delta_ang} presents this data
aggregated over all queries in each collection. We look at the angular
distances between the IEQ and each of the 90 real EQs generated for a
query, and calculate the range (i.e., the difference between the maximum
and minimum) of angular deviations from the IEQ across these 90 EQs. These
observed ranges are then averaged across all queries in each collection to
obtain the mean $\Delta\theta$ for that collection. An analogous exercise
is done to obtain mean $\Delta\mathrm{AP}$, the average of the ranges of AP
values achieved by the 90 expanded versions of each query. The aggregated
results are shown in Table~\ref{tab:delta_ang}, while
Figures~\ref{fig:ap-variation-all-qs} and~\ref{fig:angle-variation-all-qs}
visualise the distributions of AP values and angular distances from the
ideal query (\(\mathbf{Q}_{\mathrm{IEQ}}^{\mathrm{LS}} \mathrm{with}\, \lambda=1\))
across all expanded queries in the DL19-20 Passage collection.

It is clear from the table that for any given query, all the real EQs
generally lie within a relatively narrow angular region, while their
retrieval effectiveness can vary more substantially. On average, the
angular spread across all 90 variants for a query can be as low as
\(3.75^\circ\) (for the Robust04 collection, when deviations are measured
from $\mathbf{Q}_{\mathrm{IEQ}}^{\mathrm{LS}} (\lambda=0.1)$); at best, it
is \(11.15^\circ\) (for the DL19-20 Passage collection, when deviations are
measured from $\mathbf{Q}_{\mathrm{IEQ}}^{\mathrm{DFO}}$).
In contrast, the Average $\Delta\mathrm{AP}$ ranges from 0.1750 to as much
as 0.2853, indicating a broader variation in AP across expanded variants of
a query. For context, we note that the possible angular spread could, in
principle, be as much as $90^{\circ}$, for the common case where all
vectors are confined to the first quadrant; similarly, $\Delta\mathrm{AP}$
for a query could be at most 1.

\begin{table*}[h]
  \centering
  \caption{Average $\Delta\theta$ and $\Delta\mathrm{AP}$ for all collections.}
  \label{tab:delta_ang}
  \resizebox{0.9\linewidth}{!}{
    \begin{tabular}{lcccc} \toprule
      \textbf{Method}
      & \textbf{Robust}
      & \textbf{DL19-20 Passage}
      & \textbf{DL19-20 Document}
      & \textbf{DL21-22 Passage}
      \\\midrule

      \(\mathbf{Q}_{\mathrm{IEQ}}^{\mathrm{DFO}} \) 
      & \(10.02^\circ\)
      & \(11.15^\circ\) 
      & \(9.86^\circ\)
      & \(9.49^\circ\)
      \\

      \(\mathbf{Q}_{\mathrm{IEQ}}^{\mathrm{LS}} (\lambda=0.1)\)
      & \(3.75^\circ\)
      & \(5.42^\circ\)
      & \(4.04^\circ\)
      & \(4.92^\circ\)
      \\

      \(\mathbf{Q}_{\mathrm{IEQ}}^{\mathrm{LS}} (\lambda=1)\)
      & \(5.69^\circ\)
      & \(7.30^\circ\)
      & \(4.58^\circ\)
      & \(7.93^\circ\)
      \\\midrule

      Average $\Delta\mathrm{AP}$
      & 0.2215
      & 0.2853
      & 0.2624
      & 0.1750
      \\\bottomrule
    \end{tabular}
  }
\end{table*}

Thus, even though similarity to \(\mathbf{Q}_{\mathrm{IEQ}}\) is broadly
associated with higher retrieval effectiveness, the estimated correlations
are highly sensitive to small numerical variations.
In contrast, the synthetic query experiments explored a much broader range
of angular distances from \(\mathbf{Q}_{\mathrm{IEQ}}\), producing
substantial variation in Restricted AP, and yielding a clearer correlation
signal. While it is very common to use cosine similarity to quantify the
proximity between vectors, these observations suggest that it would be
worthwhile to look for more discriminatory measures that better separate
different EQs. We plan to explore this direction in future work.

\section{Limitations and future work}
\label{conclusion-future-work}
This article presents a post hoc framework for understanding the
performance of QE techniques in general, and of specific expanded queries
in particular. By constructing an ideal query for any given query, we
formalised the notion of good expansion terms and appropriate expansion
weights. We evaluated two concrete approaches for constructing ideal
queries using complete relevance assessments.
Our extensive experimental results demonstrate that the similarity between
an EQ and the ideal query is positively correlated with average
precision. Through synthetic-query analysis, we showed that retrieval
effectiveness consistently decreases as an EQ moves away from
the ideal query. We also confirmed that effective EQs clearly
separate relevant from non-relevant documents. Furthermore, across
different queries, proximity to the ideal query explains why automatic QE
techniques fail on some topics.

Thus, the ideal query serves as an effective reference point for
performance analysis of QE methods. It provides IR researchers and
practitioners working on new QE techniques with definitive answers to
questions like: what useful terms is the QE method failing to identify? are
the selected terms likely to lead to query drift? These answers, in turn,
form a basis for exploring how the proposed QE approach may be improved
(even in unsupervised settings like pseudo relevance feedback).

Nevertheless, we observe that the correlation between an EQ's similarity to
an IEQ and its AP may sometimes be only moderate. This may be attributed to
two main factors.
\begin{itemize}
\item The ideal expanded query is not necessarily unique. There may exist
  multiple, well-separated EQs that yield near-perfect retrieval
  performance. Any one of these ideal queries would serve as a useful
  target for guiding the development of practical QE methods. Efficiently
  exploring the space of possible query vectors to identify regions that
  correspond to ideal queries might constitute an interesting research
  direction. It would also be interesting to characterise queries that have
  multiple such `ideal regions', as opposed to queries that have only a
  single ideal region.
\item When measuring the similarity between real EQs and an ideal query,
  the traditional cosine formula produces a range of numerical values that
  is small, especially compared to the range of AP values corresponding to
  the same set of EQs. As a result, small random variations may
  substantially diminish the computed correlation coefficients. We propose
  to explore other similarity measures that induce a wider and more robust
  separation among the different expanded variants of a query.
\end{itemize}
Because our current framework relies on complete relevance assessments to
construct ideal queries, another important next step is learning
ideal-query representations from training data in a supervised or
semi-supervised manner. Such learned representations could guide the design
of next-generation query expansion algorithms and provide a principled
objective for optimizing retrieval models.




\bibliographystyle{ACM-Reference-Format}
\bibliography{references}

\appendix
\section{Invariance to scaling of \texorpdfstring{$\alpha$}{alpha}}
\label{sec:invariance}
To show that the AP obtained by $\mathbf{Q}_{\mathrm{IEQ}}^{\mathrm{LS}}$
(Section~\ref{ls-surrogate}) is invariant to the specific positive value
chosen for \(\alpha\), we replace \(\alpha\) with another positive constant \(\beta\).
The new target vector $\mathbf{Y}_{\mathrm{new}}$ is given by
\[ \mathbf{Y}_{\text{new}} = \frac{\beta}{\alpha}\, \mathbf{Y}_{\text{old}} = k\,
  \mathbf{Y}_{\text{old}}, \quad \text{where } k = \tfrac{\beta}{\alpha} > 0. \]
Then the least-squares solution scales accordingly:
\[ \mathbf{Q}_{\mathrm{IEQ}}^{\mathrm{LS, new}} = (\mathbf{X}^\top \mathbf{X}
  + \lambda I)^{-1} X^\top \mathbf{Y}_{\text{new}} = k\, (\mathbf{X}^\top \mathbf{X} +
  \lambda I)^{-1} \mathbf{X}^\top \mathbf{Y}_{\text{old}} = k\,
  \mathbf{Q}_{\mathrm{IEQ}}^{\mathrm{LS, old}}. \]
Consequently, the resulting BM25 score vector is also scaled by \(k\):
\[ \mathbf{X} \mathbf{Q}_{\mathrm{IEQ}}^{\mathrm{LS, new}} = k\, \mathbf{X}
  \mathbf{Q}_{\mathrm{IEQ}}^{\mathrm{LS, old}}. \]
Assume that \(\mathbf{X}\) is the full term-document matrix, instead of
comprising only the documents present in the ground truth relevance
assessment. Since multiplying by a positive constant is a strictly
monotonic transformation, the relative ordering of the document scores is
unchanged. Therefore, the retrieval ranking---and hence the AP---remains
identical. This implies that \(\alpha\) can be arbitrarily chosen (e.g., as
small as \(0.01\)) without affecting retrieval performance. After
constructing the ideal queries, we normalize them to a unit vector.

\section{AP of IEQ vs.\ number of relevant documents}
\label{additional-plots-ap-vs-relevant-docs}
\begin{figure*}[ht] 
  \centering
\includegraphics[width=0.8\linewidth]{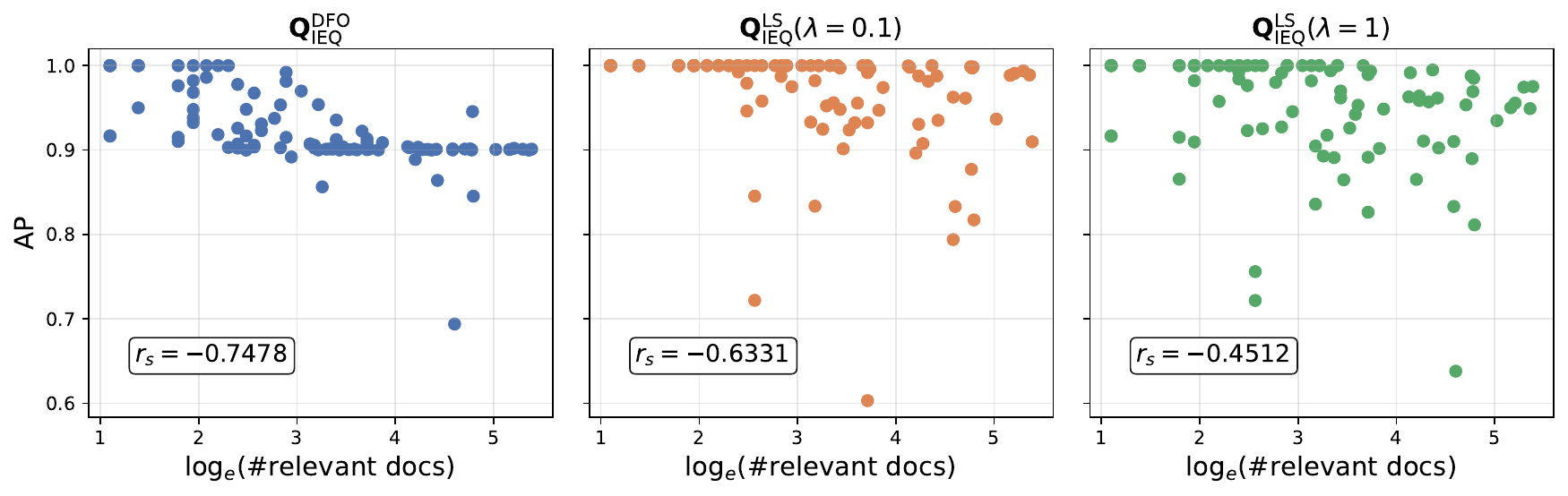}
  \caption{Per-query scatter plots of AP versus the number of relevant documents for different ideal-query formulations on the DL19-20 passage collection. The Spearman rank correlation coefficient $r_s$ between the two variables is reported in each plot.}

  \label{fig:main-ap-reldocs-dl19-20-passage}
\end{figure*} 

\begin{figure*}[ht] 
  \centering
\includegraphics[width=0.8\linewidth]{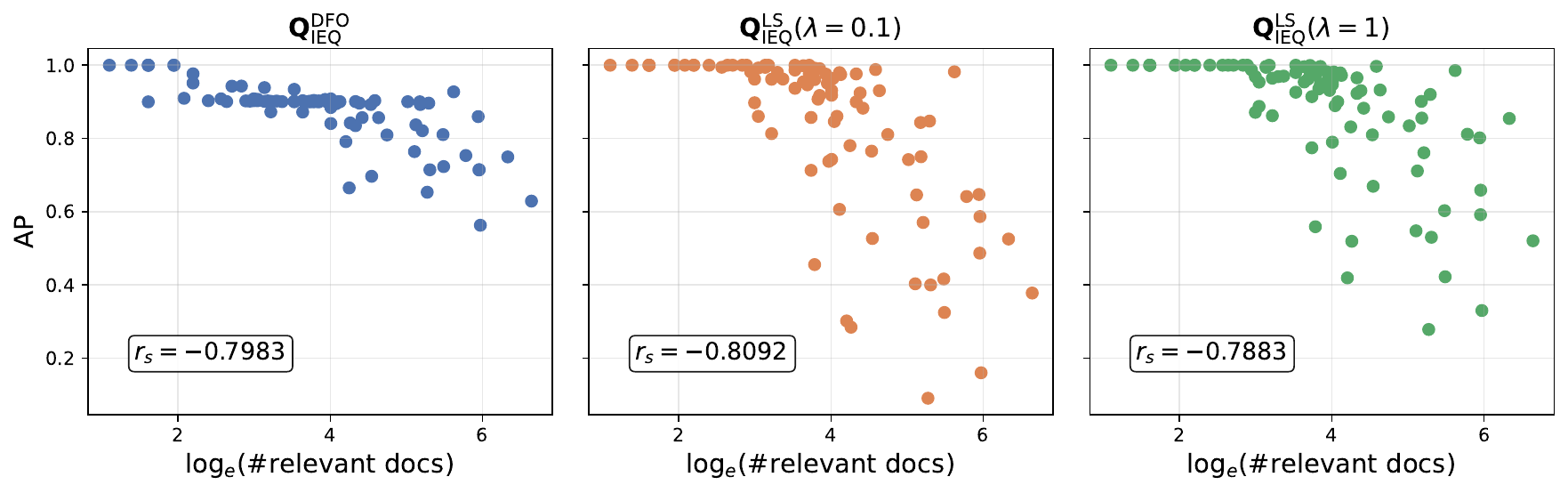}
  \caption{Per-query scatter plots of AP versus the number of relevant documents for different ideal-query formulations on the DL19-20 document collection. The Spearman rank correlation coefficient $r_s$ between the two variables is reported in each plot.}
  
  \label{fig:main-ap-reldocs-dl19-20-document}
\end{figure*}

\begin{figure*}[ht] 
  \centering
\includegraphics[width=0.8\linewidth]{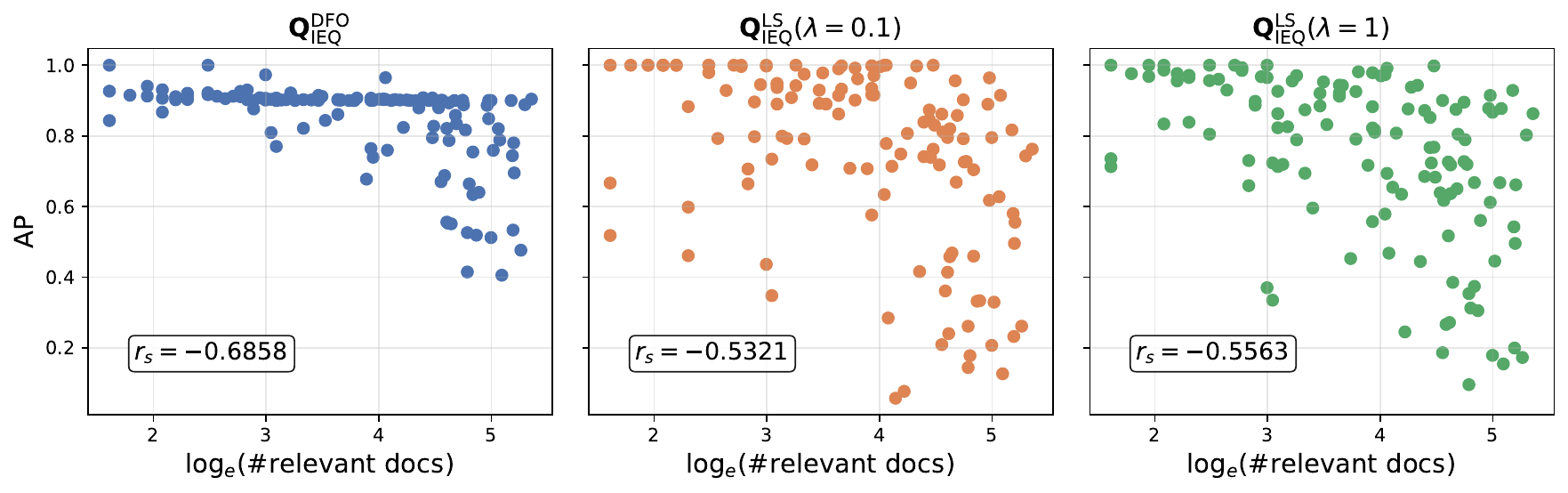}
  \caption{Per-query scatter plots of AP versus the number of relevant documents for different ideal-query formulations on the DL21-22 passage collection. The Spearman rank correlation coefficient $r_s$ between the two variables is reported in each plot. One outlier query is removed.}
  \label{fig:main-ap-reldocs-dl21-22-passage}
\end{figure*}

Figures~\ref{fig:main-ap-reldocs-dl19-20-passage},
\ref{fig:main-ap-reldocs-dl19-20-document}, and
\ref{fig:main-ap-reldocs-dl21-22-passage}, supplement
Figure~\ref{fig:main-ap-reldocs-robust} from Section~\ref{results-ieq}. The
APs achieved by ideal queries exhibit the same trend across collections:
the number of relevant documents is strongly negatively correlated with AP.
This phenomenon is particularly pronounced for DFO-based ideal queries, and
is generally more prominent for the document collections than for the
passage collections.

\section{Choice of \texorpdfstring{$\lambda$}{lambda}}
\label{ablation-studies}
\begin{figure}[b]
    \centering
    \includegraphics[width=0.5\linewidth]{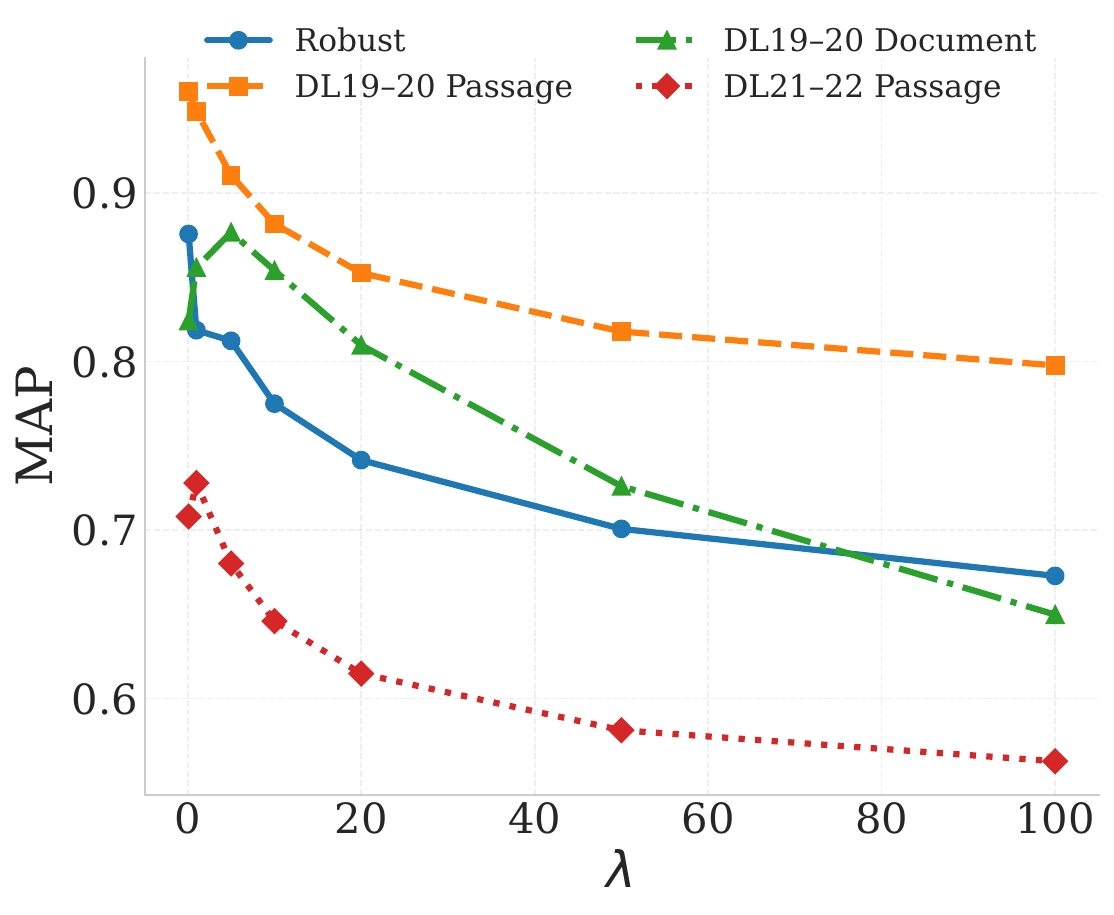}
    \caption{Relationship between the MAP values and $\lambda$ for LS based ideal queries.}
    \Description[map vs lambda plot]{Relationship between the MAP values and $\lambda$ for LS based ideal queries.}
\label{fig:lambda_vs_map}
\end{figure}

In Section~\ref{results}, for the results pertaining to ideal queries
constructed using the least-squares (LS) method, we used $\lambda = 1.0$
and 0.1 as the regularization parameter. In this appendix, we take a more
detailed look at the effect of the value of $\lambda$.
Figure~\ref{fig:lambda_vs_map} shows (for all collections) how MAP varies
as \(\lambda\) increases. MAP is highest for relatively small values of
\(\lambda\) (typically between 0 and 1). From
Equation~\eqref{eq:ls-optimization}, it is clear that for higher values of
\(\lambda\), progressively less importance is attached to the actual
similarity values of relevant and non-relevant documents, and more
importance is attached to just $\|\mathbf{Q}\|$. Naturally, this leads to a
gradual decline in retrieval effectiveness.

\begin{figure}[t]
    \centering
    \includegraphics[width=0.5\linewidth]{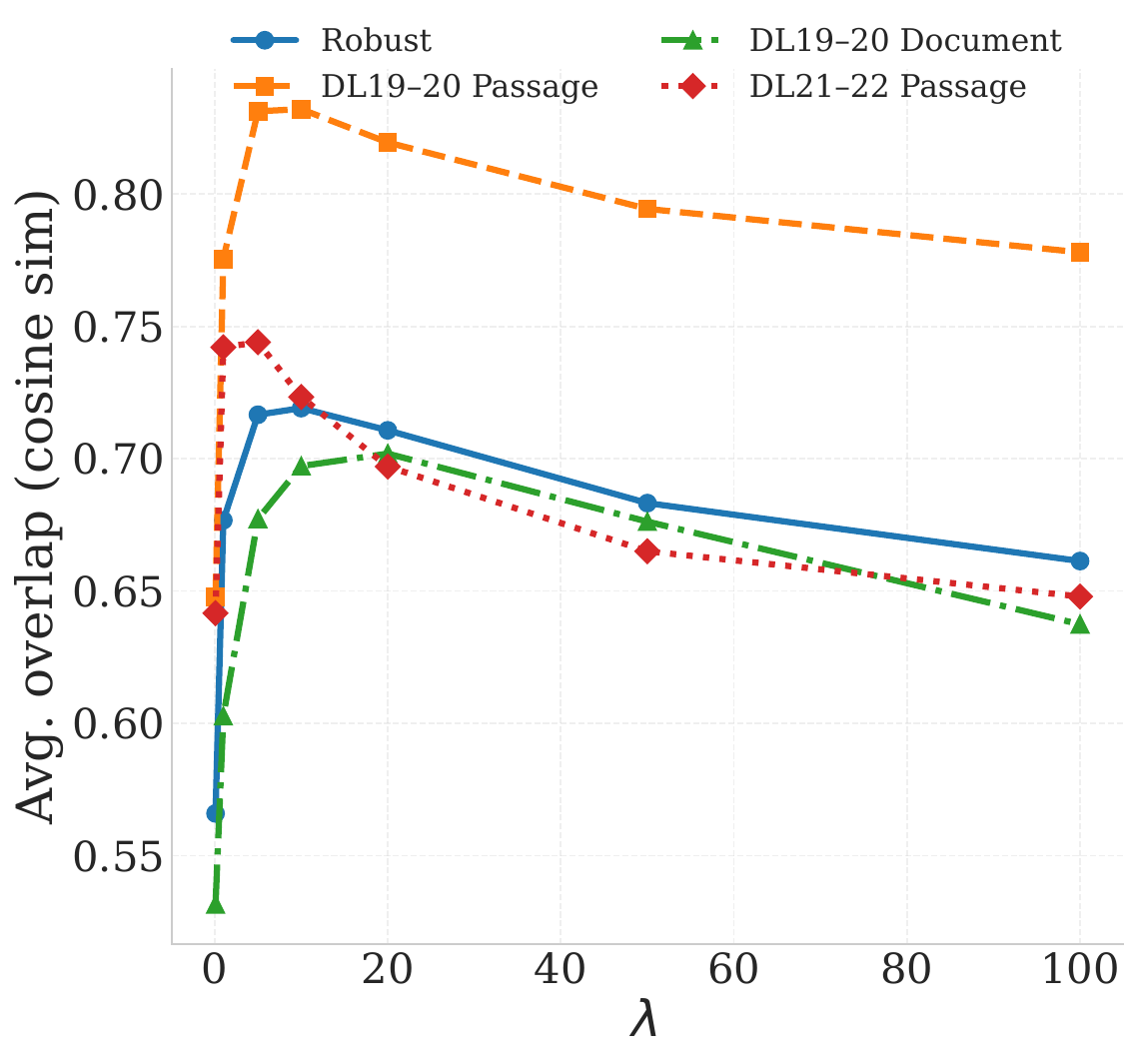}
    \caption{Relationship between the average overlap with DFO based ideal query the LS based ideal queries generated with different $\lambda$.}
    \Description[avg overlap vs lambda plot]{Relationship between the MAP values and $\lambda$ for LS based ideal queries.}
\label{fig:lambda_vs_avg_overlap}
\end{figure}

We also study the effect of different values of \(\lambda\) on the average cosine
similarity between a DFO-based ideal query
$\mathbf{Q}_{\mathrm{IEQ}}^{\mathrm{DFO}}$ and the corresponding LS-based
ideal query $\mathbf{Q}_{\mathrm{IEQ}}^{\mathrm{LS}}$.
Figure~\ref{fig:lambda_vs_avg_overlap} reports, for each collection, the similarity between these two ideal queries, averaged across all queries.
Across all collections, the average overlap initially increases with
\(\lambda\), reaches a maximum, and then gradually declines. The maximum
overlap is attained for \(\lambda\) values between 1 and 5 for all
collections. 
These results indicate that moderate regularization produces LS-based
ideal queries that are most similar to the DFO-based formulation.

\section{Variability across queries}
\label{variability}
Before studying the variability across different user queries, we first
compare it to the variability across different expanded variants of a
single query. For each query $Q$, we consider the $k=90$ different expanded
versions, and compute both the mean ($\overbar{\theta_{Q}}$) and the
variance ($\mathit{Var}({\theta_{Q}})$) of their angular distances from the
corresponding ideal query. We refer to the variance of the mean values
($\overbar{\theta_{Q}}$) across queries as \emph{between-query variance}
(BQV), while we use the term \emph{within-query variance} (WQV) to refer to
the mean of the variances ($\mathit{Var}({\theta_{Q}})$). The same process
is repeated for AP instead of angular distance. The resulting BQV/WQV
ratios are reported in Table~\ref{tab:ang-bcv-wcv}. A larger BCV/WCV ratio
indicates that the variation across queries is greater than the variation
among different query expansion methods for the same query. As shown in
Table~\ref{tab:ang-bcv-wcv}, the BQV/WQV ratios for angular distance range
from 6.42 to 17.74 across the four collections and ideal query
representations, indicating substantially greater variation across queries
than among QE methods for the same query. A similar pattern is observed for
AP, where the BQV/WQV ratios range from 6.76 to 8.66. Together, these
results suggest that query-specific differences contribute substantially
more to the observed variation than the choice of QE method.

\begin{table*}[h]
  \centering
  \caption{BQV/WQV ratios for both angular distances and AP for all
    collections. Larger BQV/WQV ratios indicate that the separation between
    query groups is substantially greater than the variability within each
    group.}
  \label{tab:ang-bcv-wcv}
  \resizebox{1\linewidth}{!}{
    \begin{tabular}{lcccc} \toprule

      \textbf{Method}
      & \textbf{Robust}
      & \textbf{DL19-20 Passage}
      & \textbf{DL19-20 Document}
      & \textbf{DL21-22 Passage}
      \\\midrule

      \(\mathbf{Q}_{\mathrm{IEQ}}^{\mathrm{DFO}} \) 
      & 15.79
      & 15.95 
      & 17.74 
      & 9.65
      \\

      \(\mathbf{Q}_{\mathrm{IEQ}}^{\mathrm{LS}} (\lambda=0.1)\)
      & 14.37
      & 6.42
      & 9.20
      & 10.38
      \\

      \(\mathbf{Q}_{\mathrm{IEQ}}^{\mathrm{LS}} (\lambda=1)\)
      & 11.44
      & 8.89
      & 9.13
      & 8.76
      \\\midrule

      AP 
      & 8.66 
      & 7.07 
      & 6.76 
      & 6.85 \\\bottomrule
    \end{tabular}
  }
\end{table*}


\end{document}